\documentclass{ws-procs9x6}
\newcommand{\be}{\begin{eqnarray}}
\newcommand{\ee}{\end{eqnarray}}

\begin{document}

\title{TASI 2009 Lectures: \\ Searching for Unexpected Physics at the LHC}

\author{Kathryn M. Zurek}

\address{Department of Physics, University of Michigan\\
Ann Arbor, MI 48109 USA\\
E-mail: kzurek@umich.edu}

\begin{abstract}

These TASI lectures consider low mass hidden sectors from Hidden Valleys, Quirks and Unparticles.  We show how each corresponds to a different limit of the same class of models: hidden sectors with non-abelian gauge groups with mass gaps well below a TeV that communicate to the Standard Model through weak scale suppressed higher dimension operators.  We provide concrete examples of such models and discuss LHC signatures.  Lastly we turn to discussing the application of Hidden Valleys to dark matter sectors.

\end{abstract}


\bodymatter

\section{Introduction}\label{aba:sec1}

I have been given the task of lecturing on ``unexpected physics at the LHC," and let me begin with a comment on the irony of the title.  Because of course any subject which warrants a series of TASI lectures is not totally unexpected physics.  

So what do we mean by the words ``unexpected'' signals at the LHC?  Most of the effort for searches of physics beyond the Standard Model has centered on solutions to the so-called hierarchy problem, which reduces to the question of ``why is the Higgs boson so light?''  Because naturally one would expect, without an inordinate amount of fine-tuning, that the Higgs boson would receive radiative corrections that push its mass up to the Planck scale.  We as a particle physics community have largely focused our efforts on solving this problem by adding new dynamics at the TeV scale.   The most popular types of this type are
\begin{itemize}
\item Supersymmetry
\item Extra dimensions (large, warped, Higgsless)
\item Technicolor
\item Little Higgs.
\end{itemize}
And the list goes on.  We've discovered that the last three are intimately connected to each other by dualities, so really solutions to the hierarchy problem can be be termed of the supersymmetric type or the strong dynamics type.

Most of the phenomenology and search techniques employed have focused on such models which solve the hierarchy problem.  These models have many features to distinguish them one from the other. However they do share a few things in common: new states at the TeV scale which couple through weak or strong interactions to Standard Model (SM) states.  
 
 The focus has been on the continued search for such heavy states.  Now as time has gone along, people have not discussed only solutions to the hierarchy problem; they have also looked at other new heavy states at the TeV scale, such as
 \begin{itemize}
 \item $Z'$
 \item Fourth generation
 \item Leptoquarks
 \item Color octet.
\end{itemize}
 And again the list goes on.  Why look at such things?  They might be there.  Nature is not simple.  But here again, the focus has been on new states residing at the TeV scale.
 
 The moral of the story here is that theorists and experimentalists alike have been focused on the search for new heavy objects, and the focus has been on pushing to higher energies in order to access those heavier states.
 
\begin{figure}
\begin{center}
\psfig{file=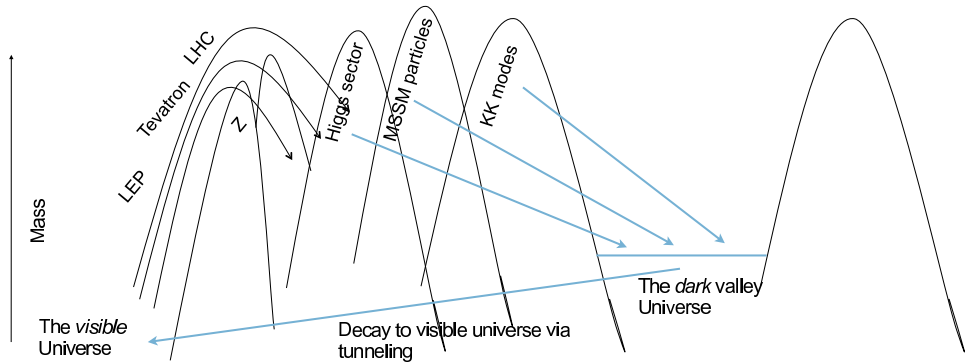,width=5.0in}
\end{center}
\caption{A depiction of a Hidden Valley.  The peaks represent massive states which may connect the Standard Model sector to light states in the hidden sector.}
\label{SUSYValley}
\end{figure}

 Now we are in a position to answer the question I posed initially: what do we mean by unexpected physics at the LHC?  The focus of these lectures is on classes which escape the traditional search techniques in many cases because {\em they feature new low mass states} in a hidden sector.  Such low mass states could have escaped detection particularly when new heavy states must be produced which then decay into lighter states.  A good visual picture of this scenario can be seen in Fig.~(\ref{SUSYValley}), and it was this type of picture Matt Strassler and I had in mind when we developed Hidden Valleys \cite{HV}.  While this is one class of models which gives rise to unexpected signatures at the LHC, it is not the only one.   We will focus on three classes of models which generate related phenomenology at the LHC: Hidden Valleys (HV), Quirks \cite{quirks} and Unparticles \cite{unparticle}.  We will also see that these classes of models have potentially significant implications for dark matter searches.
 
\begin{figure}
\begin{center}
\psfig{file=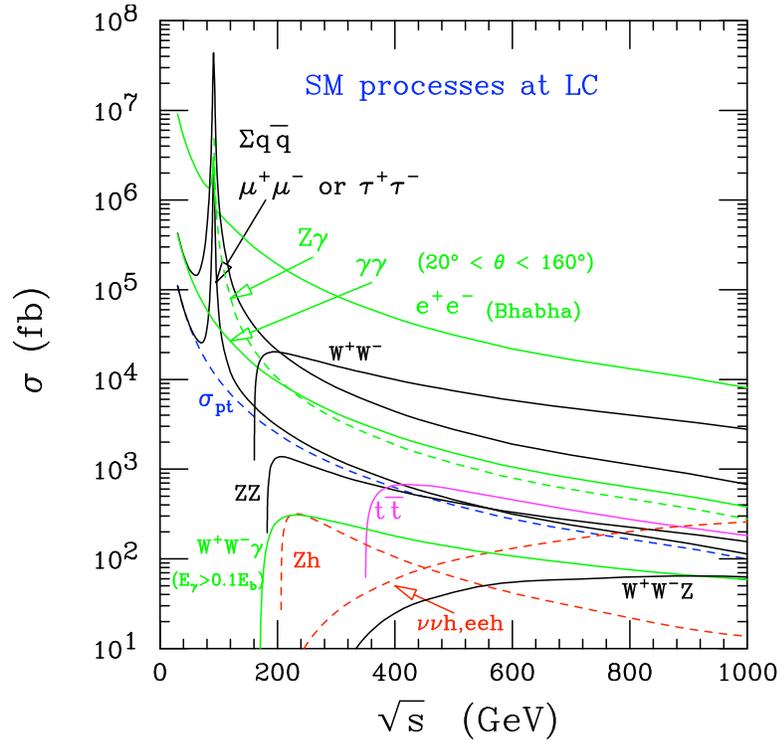,width=4.0in}
\end{center}
\caption{Beating backgrounds at high center of mass energies, from T. Han's TASI lectures \cite{THanTASI}.}
\label{Backgrounds}
\end{figure}

Thus the focus of these lectures will be "unexpected" physics from hidden sectors with low mass states. Searching for such low mass sectors at a high energy collider such as the LHC could be difficult because of the very large backgrounds associated with production of soft low mass particles in high multiplicities.  Thus in many cases such sectors could have escaped detection, and will continue to escape detection unless new searches are designed to look for them.  Typically background removal centers on high $p_T$ objects, and cuts on high invariant mass of objects in final states.   The reason for doing this is clear, as shown in Fig.~(\ref{Backgrounds}).  Many backgrounds fall off rapidly at high center of mass energy.  On the other hand, resonant production of new states at high center of mass energy enhances their production and makes such signals visible over large SM backgrounds.  That is, new physics does not drop as quickly with $p_T$ cuts because heavy objects are being created.  Other variables (such as invariant mass) create additional handles.

We now turn to describing in detail some examples of these low mass hidden sectors.  I should warn at the outset that these lectures will not contain exhaustive referencing, but only a few papers that are directly utilized in these lectures.
 
 \section{Types of Unexpected Physics at the LHC}
 
 In the introduction we defined the types of models at the LHC which we will focus on as being ``unexpected'': new physics which could be missed by the focus on heavy objects and high $p_T$ final states.   
 
 These models consist of a new, low mass hidden sector which connects to the standard model either through heavy states ({\em i.e. } higher dimension operators), or through kinetic mixing.  The dynamics in the hidden sector may be complex with new ``dark'' forces which couple only to states in the hidden sector, as well as complex mass patterns and cascade decays within the hidden sector.  The common features between these models are
 \begin{itemize}
 \item a hidden sector which is SM neutral, and
 \item a connector sector which is charged under both the standard model and the hidden sector. 
 \end{itemize}
 The heavy connectors are represented in Fig.~(\ref{SUSYValley}) as peaks, and the light hidden sector is represented as a valley.
 
The connector sectors could be many things, including many of the new heavy states that we discussed in the introduction, such as
 \begin{itemize}
 \item Supersymmetric states
 \item $Z'$
 \item Higgs
 \item Fourth generation.
 \end{itemize}
 The list could be as long as your creativity allows for.
 
\begin{figure}
\begin{center}
\psfig{file=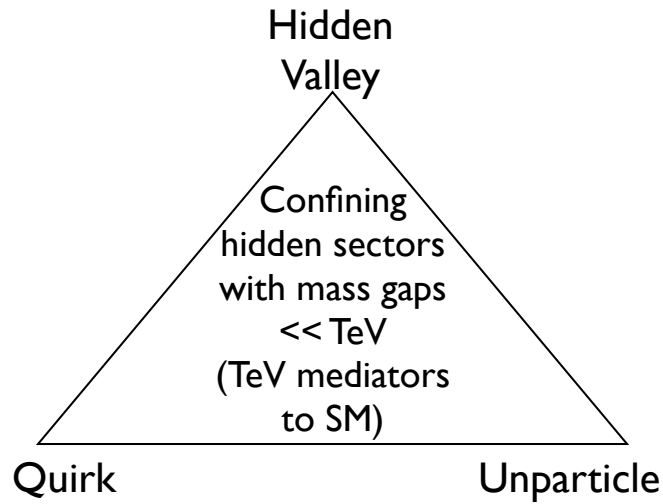,width=3.5in}
\end{center}
\caption{Three types of related models which are the subject of these lectures.}
\label{HVtriangle}
\end{figure}

\begin{figure}
\begin{center}
\psfig{file=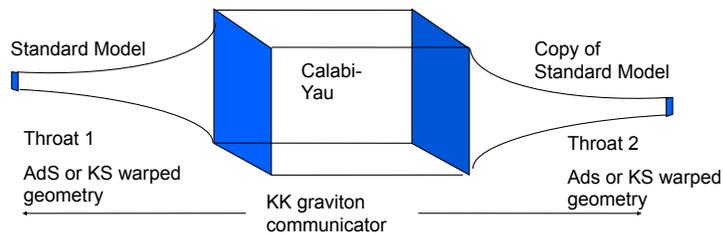,width=4.0in}
\end{center}
\caption{A schematic of string-motivation for Hidden Valleys, where the SM resides in one throat, and the hidden sector resides in the second throat.}
\label{HVthroat}
\end{figure}

 As for the content of the hidden sector, in the first two parts of this three part lecture series, we focus on models where the hidden sector is characterized by strong dynamics.  The three types of models we consider are illustrated in Fig.~(\ref{HVtriangle}).  But as we know from the AdS/CFT correspondence, in certain limits QCD-like (conformal in the ultraviolet) theories are dual to a warped extra dimension.  Thus the interest of these models can be extended to models which are string motivated, for example a hidden sector in an extra dimensional warped throat with some cut-off at the tip of the throat, as illustrated schematically in Fig.~(\ref{HVthroat}).    To be more concrete, the types of sectors that could reside in the Hidden Valley are
 \begin{itemize}
 \item QCD-like theory with $F$ flavors, $N$ colors
 \item QCD-like theory with only heavy quarks (Quirk limit)
 \item Pure glue theory
 \item {\cal N} = 4 SUSY conformal
 \item Randall-Sundrum (RS) or Klebanov-Strassler (KS) throat (with Seiberg duality cascade in KS throat)
 \item  Partially Higgsed $SU(N)$
 \item Banks-Zaks infrared fixed point (Unparticle limit)
 \end{itemize}
 I have noted the various limits (quirk and unparticle) of the Hidden Valley that we will discuss, as illustrated in Fig.~(\ref{HVtriangle}).  Note however that much of the possible range of models which could be studied has not been: this is an area that is still relatively little explored.

 A particularly nice way of organizing these sectors in terms of $\alpha$, the gauge coupling, and $\beta$, the running of the coupling, which I summarize for convenience in Table~(\ref{MS}) \cite{refMS}.  One can see that all these hidden sectors with strong or quasi-conformal dynamics can be thus classified for reference as we discuss various types .
 
 \begin{table}
{\begin{tabular}{l|lll}
 & zero $\beta$ (CFT) & small $\beta$ & large $\beta$ \\\colrule
small $\alpha N$ & Banks-Zaks & Perturbed Banks-Zaks & \\
& ${\cal N} = 4$ SUSY & unparticle with mass gaps & won't last \\
& unparticles & technicolor UV & \\  \\
large $\alpha N$ & ${\cal N} = 4$ SUSY & Perturbed Seiberg CFT & QCD IR \\
& Generic Seiberg CFT & & Infracolor IR \\ \\
extreme $\alpha N$ & RS bulk & deformed RS bulk & RS IR brane \\
& & KS bulk & \\\colrule
\end{tabular}}\label{MS}
\end{table}
 
 The introduction of these low mass hidden sectors has also resulted in much fruitful thinking about low mass dark matter sectors, and their connection to cosmology.   We will return to discussing this topic in the third lecture.  It is particularly timely given the hints for signals in the dark sectors.  But next we turn to discussing particular examples of these low mass hidden sector models.
 
Before we go on to examples however, we want to emphasize a few things.  First, the phenomenology of the hidden valleys will be strongly determined by the number of hidden sector $v$-quarks and whether they are lighter or heavier than the confinement scale of the strong gauge group in the hidden sector.  The canonical Hidden Valley case will have only one or two quarks lighter than the confinement scale in the hidden sector.  The quirk limit will have only quarks heavier than the confinement scale.  As we will see, this leads to very different phenomenology.  The unparticle limit will correspond to a theory with conformal behavior. We now turn to examples.
 
 \section{An example of a Hidden Valley}
 
 To gain some intuition about how these confining hidden sector models work, we consider a simple QCD-like hidden sector, having only two light flavors. Light or heavy flavor is defined with respect to the confinement scale: light quarks have masses $m_v < \hat{\Lambda}$ and heavy quarks have masses $m_v > \hat{\Lambda}$.  As we will see, the number of light or heavy quarks will have a lot to do with the phenomenology of the hidden sector.
 
The hidden sector quarks we will call $v$-quarks ($v$ for valley), and for two light flavors we will give them the labels $v_1$ and $v_2$.  For further concreteness, the connector of this hidden sector to the SM will be a $Z'$, whose charges are from a $U(1)_\chi$ gauge group.  One could have just as well chosen a hidden Higgs boson.  The $Z'$ will have a TeV mass, and both the $v$-quarks and the confinement scale will have a much lower mass.  The $Z'$ from the $U(1)_\chi$ is a convenient choice because we know how to arrange things such that the SM plus hidden sector is anomaly free.  The charges are shown in Table~(2).
 
 \begin{table}
{\begin{tabular}[c]{||c||c|c|c|c|c|c||c|c|c|c||c|c||}\hline
\ & $q_i$ &$\bar u_i$& $\bar d_i$ 
& $\ell_i$ & $e^+_i$ & $N_i$ & $U$ & $\bar U$ & $C$ & $\bar C$ & $H$ & $\phi$
\\ \hline 
$SU(3)$ & ${\bf 3}$ & ${\bf \overline 3}$  & ${\bf \overline 3}$ 
&  ${\bf 1}$  & ${\bf  1}$ &  ${\bf 1}$  & ${\bf  1}$ 
&  ${\bf 1}$  & ${\bf  1}$ &  ${\bf 1}$  & ${\bf  1}$ &  ${\bf 1}$  
\\ \hline
$SU(2)$ & ${\bf 2}$ & ${\bf 1}$  & ${\bf  1}$ 
&  ${\bf 2}$  & ${\bf  1}$ &  ${\bf 1}$  & ${\bf  1}$ 
&  ${\bf 1}$  & ${\bf  1}$ &  ${\bf 1}$  & ${\bf  2}$ &  ${\bf 1}$  
\\ \hline
$U(1)_Y$ & $\frac16$ 
& $-\frac23$  & $\frac13$ 
& $-\frac12$ & $1$  & $0$ & $0$ & $0$ & $0$ & $0$ & $\frac12$ & $0$ 
 \\  \hline 
$U(1)_\chi$ & $-\frac15$ & $-\frac15$  & $\frac35$ 
& $\frac35$ & $-\frac15$ & $-1$ 
& $q_+$ & $q_-$ & $-q_+$ & $-q_-$ 
& $\frac25$ & $2$
\\ \hline
$SU(\hat{N})$ & ${\bf 1}$ & ${\bf 1}$  & ${\bf  1}$ 
&  ${\bf 1}$  & ${\bf  1}$ &  ${\bf 1}$  
&  ${\bf \hat{N}}$  & ${\bf  \overline{\hat{N}}}$ 
&  ${\bf \hat{N}}$  & ${\bf  \overline{\hat{N}}}$ 
& ${\bf  1}$ &  ${\bf 1}$  
\\ \hline
\end{tabular}}
\label{charges}
\end{table}
 
\begin{figure}
\begin{tabular}{cc}
\psfig{file=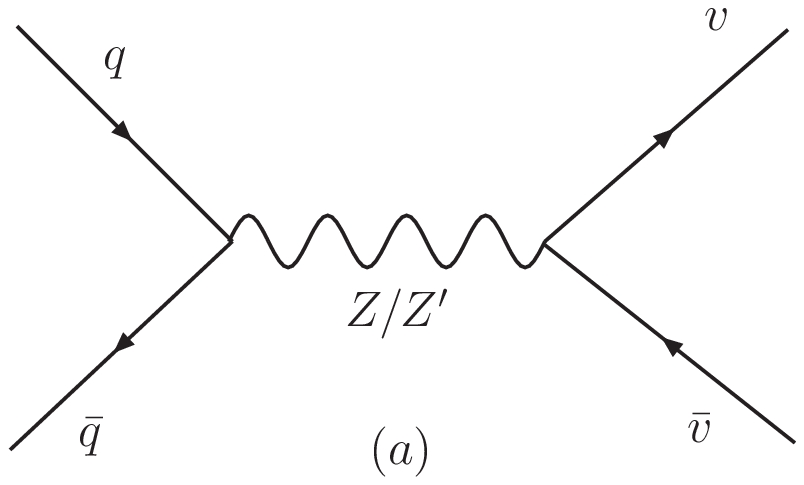,width=2.0in} & \psfig{file=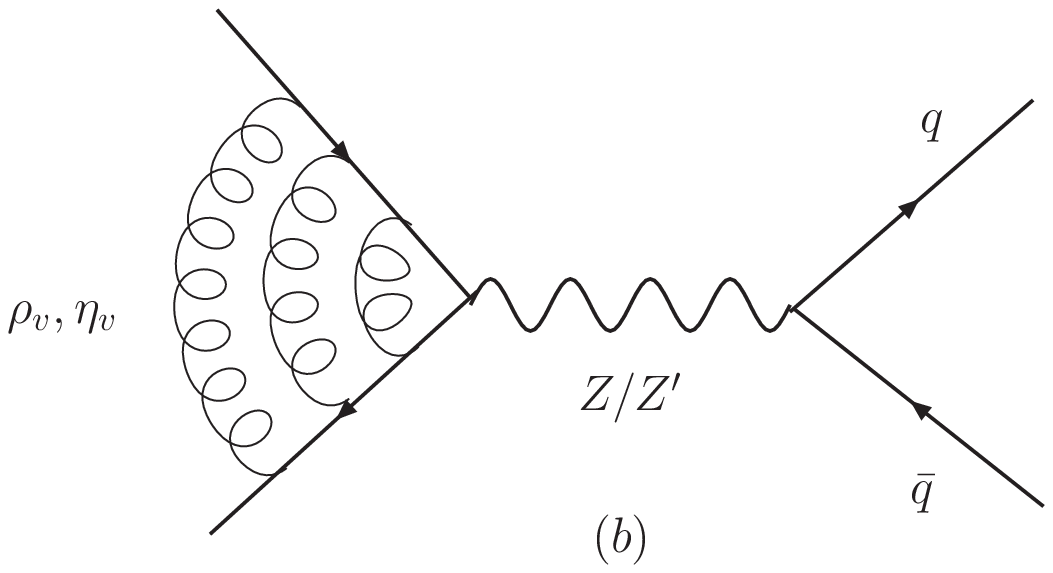,width=2.0in}
\end{tabular}
\caption{a) Production of $v$ quarks through the connector $Z'$.  b) Decay of the HV $v$-hadrons.  In a one light flavor model, the relevant asymptotic degrees of freedom are the $\rho_v$ and $\eta_v$, analogues of SM $\rho$ and $\eta$.  The pseudoscalar $\eta_v$ prefers to decay to the heaviest flavor available, whereas the vector $\rho_v$ decays democratically.}
\label{feyna}
\end{figure}

 With these charges, production and decay processes at hadron colliders are shown in Fig.~(\ref{feyna}).  Fig.~(\ref{feyna})b shows the decay of the vector and pseudo-scalar states $\rho_v$ and $\eta_v$ which are the analogue of SM $\rho$ and $\eta$.  The $\rho$ and the $\eta$ are the asymptotic states in a one light flavor model, whereas pions are the asymptotic states in a two light flavor model.   The $v$-pions are are linear combinations of the two $v$-quarks in a two light flavor model,
 \begin{eqnarray}
 \bar{v}_1 v_2 \Leftrightarrow \pi_v^+ \\ \nonumber
 \bar{v}_1 v_1+\bar{v}_2 v_2 \Leftrightarrow \pi_v^0 \\
 \bar{v}_2 v_1 \Leftrightarrow \pi_v^- \nonumber.
 \end{eqnarray}
We note that the $v$-pions (even those labelled $+$ and $-$) are electrically neutral--the labels are simply meant to elucidate the analogy with SM pions.  In analogy with the SM, the $v$-pion masses are
 \begin{equation}
 m_{\pi_v}^2 \simeq \hat{\Lambda} m_v,
 \end{equation}
 where $\hat{\Lambda}$ is the confinement scale of the hidden strong group.
 
\begin{figure}
\begin{center}
\psfig{file=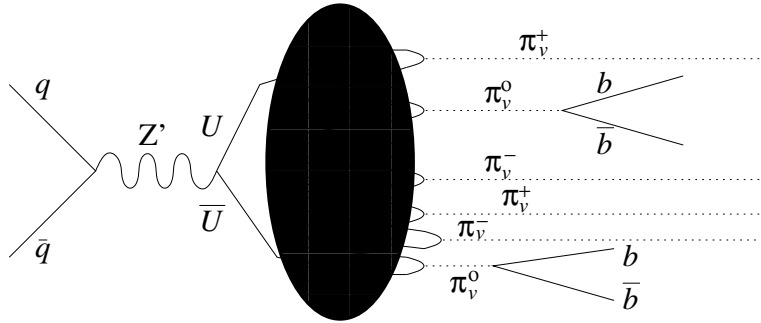,width=4.0in}
\end{center}
\caption{The production and hadronization of $v$-quarks (labeled by $U$ in the figure).  For two light flavor models, the asymptotic degrees of freedom, $v$-pions, are either stable, or decay to the heaviest flavor kinematically available.}
\label{2LF}
\end{figure}

 In each event many $v$-pions will be produced.  A typical event for a two light flavor model is shown in Fig.~(\ref{2LF}).  Once the $v$-pions are produced they can decay back through the heavy $Z'$ to SM states, as shown in Fig.~(\ref{feyna})b.  Just as SM pions preferentially decay to heavy SM states, so the $v$-pions also preferentially decay to the heaviest SM pair kinematically available.  
 
 Now because the $v$-pions are light, the decay through the heavy $Z'$ can be suppressed, and the resulting lifetimes can be long.  It turns out when calculated explicitly, we have for the $v$-pion \cite{HV}
 \begin{equation}
 \Gamma_{\pi_v \rightarrow b \bar{b}}\simeq 6 \times 10^9 \mbox{ sec}^{-1} \frac{f_{\pi_v}^2 m_{\pi_v}^5}{20 \mbox{ GeV}^7} \left(\frac{10 \mbox{ TeV}}{m_{Z'}/g'}\right)^4,
 \end{equation}
 where the hidden pion scale is $f_{\pi_v} \simeq \hat{\Lambda}$.  One sees that for $v$-pion masses much below 20 GeV, $v$-pion lifetimes are long enough to result in macroscopic decay lengths, so that a displaced vertex could appear in the detector.  Such displaced vertices have become an increasing focus of experimental searches.
 
\begin{figure}
\begin{center}
\psfig{file=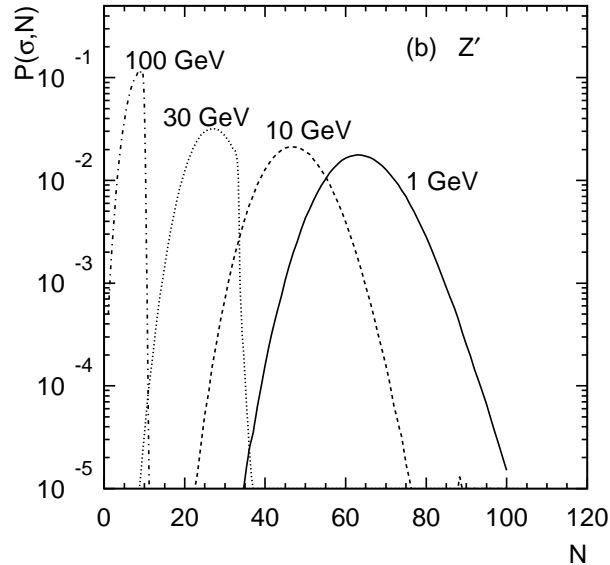,width=3.5in}
\end{center}
\caption{Differential cross-section distribution of $v$-hadron multiplicity, for various $v$-hadron masses.  These $v$-hadrons are produced from the decay of an on-shell 1 TeV $Z'$ connector.}
\label{multiplicity}
\end{figure}

 The multiplicity of $v$-pions in an events will depend both on the confinement scale and the center of mass energy of the event, but roughly the scaling is 
 \begin{equation}
 N_{\pi_v} \sim E_{cm}/m_{\pi_v}.
 \end{equation}
For concreteness, we show in Fig.~(\ref{multiplicity}) the results from simulating the hadronization for the case that the two $v$-quarks are produced through an intermediate on-shell $Z'$ \cite{HVpheno}.  
 
 The other important feature to note in these models is that the isospin +1 and -1, $\pi_v^+$ and $\pi_v^-$ states are stable: this combination of $v$-quarks does not couple to the connector $Z'$.  One might worry that there is a cosmological issue with the $v$-pions in this case.  This is no problem, however, since the $\pi_v^\pm$ are typically somewhat heavier than the $\pi_v^0$, so that they rapidly annihilate in the early universe to $\pi_v^0$ 's which then decay through the $Z'$.  Thus there is little relic abundance of $\pi_v^\pm$.  On the other hand, these isospin $\pm 1$ states will still give rise to a large missing energy signal at the LHC.

This is just one simple model with two light flavors.    As we alluded to earlier, other simple variants can be constructed.  For example, simply by positing one light flavor instead of two, the phenomenology becomes very different, and in fact much simpler to extract from the data at the collider.  In the case of one light flavor, the light degrees of freedom, the $\eta_v$ (pseudoscalar) and $\rho_v$ (vector) have masses
 \begin{equation}
 m_{\rho,\eta} \simeq \hat{\Lambda}
 \end{equation}
 and not the geometric average of the confinement scale and $v$-quark mass.
 For collider phenomenology, the important point is that while the pseudoscalar will still decay predominantly to heavy flavor, the vector will have {\em democratic} decays to all flavors.  As a result, it may be possible to tag such events using multiple leptons (especially muons) from the decay of the vector.  This greatly increases the ease with which these events can be separated from backgrounds.  For the model of Table~(\ref{charges}) (with only one of the flavors taken to be light), one finds the branching fraction to muons, for example, is approximately 4\% \cite{HVpheno}.
 
 To see a little more systematically how signal and background separation might happen, we return to the issues of triggering.  Backgrounds can be removed by triggering on hard objects.  Muons are especially clean: they live long enough to reach muon chambers where their properties can be very precisely extracted.  However, electrons and hard jets can also be efficient tools, though we focus on the muons in these events as handles.  
 
 Relative to more ``expected'' signals from new heavy physics, there is a greater challenge in searching for HVs because the high multiplicity of $v$-hadrons means that the center of mass energy is divided among many objects, which are as a result typically much softer.  So for jets resulting from $v$-hadron decay, there will be larger QCD backgrounds.  This makes detection especially difficult in the absence of displaced vertices, which might be used as a handle to reduce the QCD backgrounds.   A investigation has been carried out on how to search for hidden valleys in the absence of clean lepton handles \cite{HVheavy}, but we will not discuss this direction further in this lecture.  Backgrounds are daunting, and such a search will be difficult.  
 
\begin{figure}
\begin{tabular}{cc}
\psfig{file=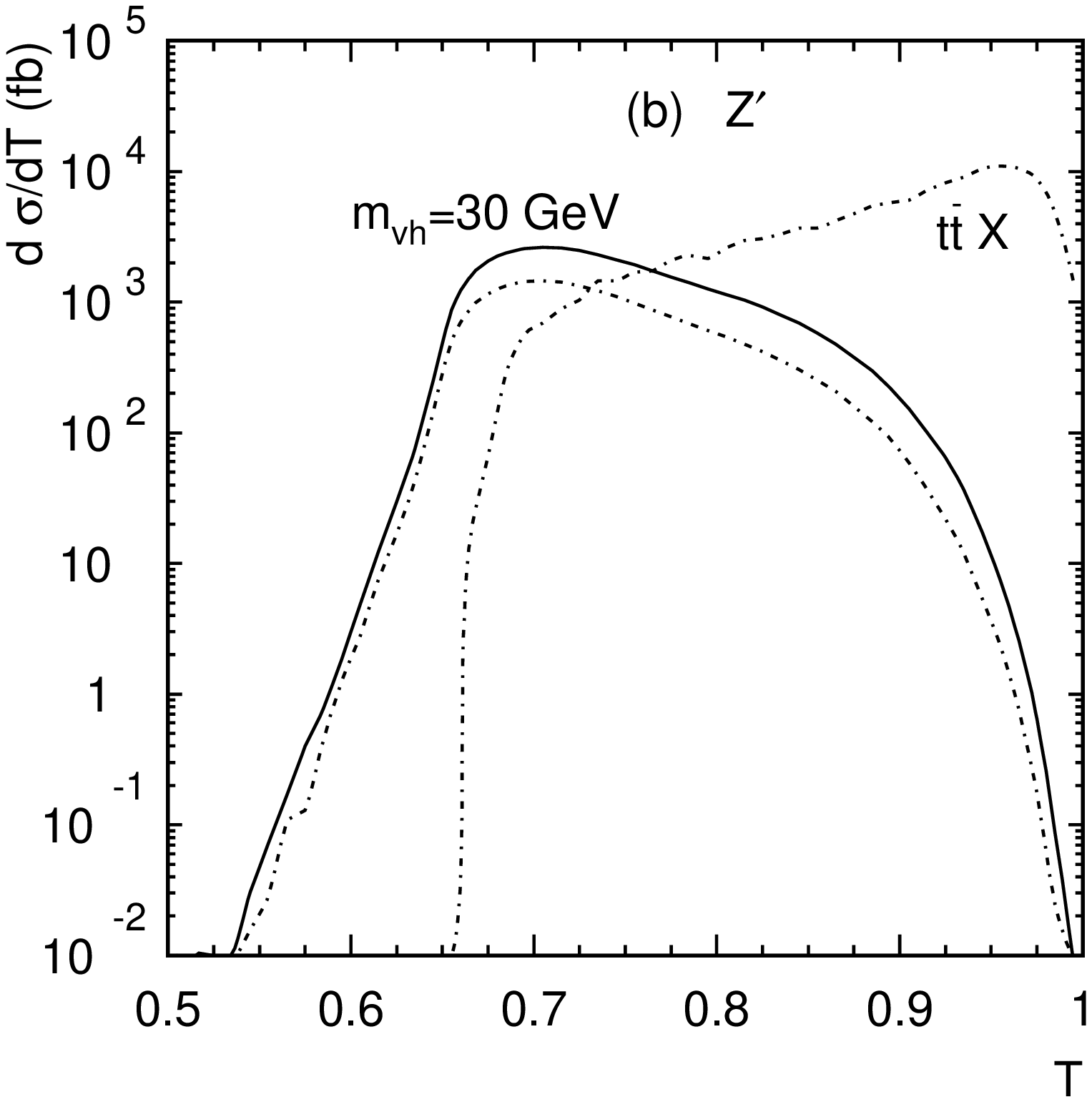,width=2.5in} & \psfig{file=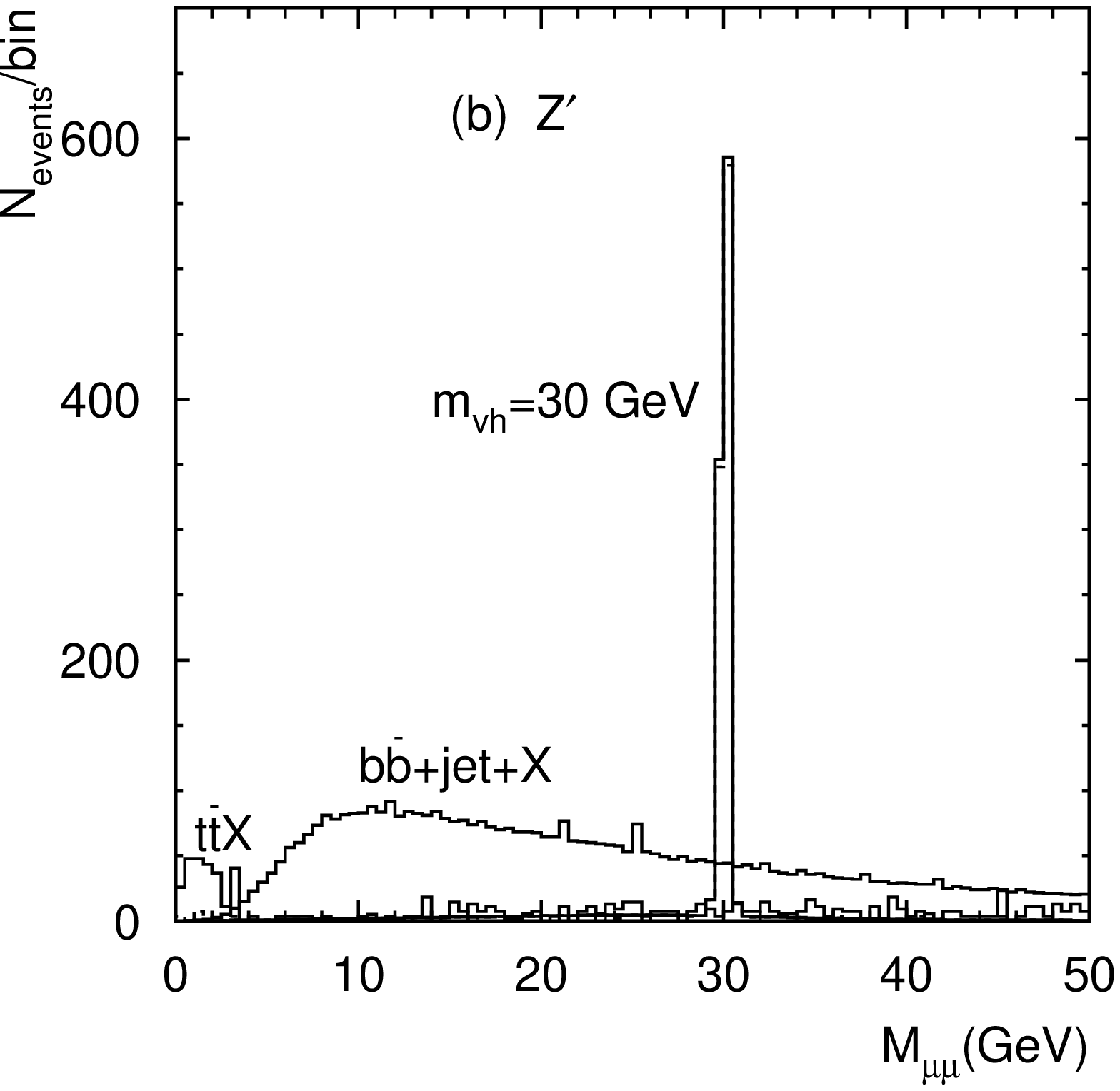,width=2.5in}
\end{tabular}
\caption{a) Thrust distribution of decay products of $v$-hadrons produced from decay of on-shell 1 TeV $Z'$ as compared to $t\bar{t}$ background.  The HV events are rounder than Standard Model backgrounds. b)  Invariant mass distribution of muon pairs from $v$-hadron decays.  The muon pairs can be used to reconstruct $v$-hadrons and separate them from Standard Model backgrounds \cite{HVpheno}.}
\label{HVpheno}
\end{figure}

 However, in the one light flavor case, there is a 4\% branching fraction to muons.  The muons can be used to reconstruct the low mass resonances which efficiently eliminates the backgrounds.  We discuss this case now in more detail.  The set of cuts one designs to eliminate the backgrounds for this particular case can be summarized as follows \cite{HVpheno}
 \begin{itemize}
 \item HV events occur at high center of mass energy, since most are produced through an on-shell $Z'$.  The high center of mass can be used as a cut to eliminate soft SM backgrounds.
 \item HV events are typically more spherical than the SM background, as shown in Fig.~(\ref{HVpheno}a).
 \item HV events have {\em very narrow} low mass resonances which reconstruct to the $v$-hadron mass.  This cut is most efficient of all for eliminating the SM backgrounds, as shown in Fig.~(\ref{HVpheno}b).
 \end{itemize}
 This is not meant to be an exhaustive description, but only a summary of the types of searches one could design to look for HVs.   Of course, what we have illustrated is how one can go about designing a search for this type of HV where there are light vector resonances which decay to muons some fraction of the time.  It can be research for you to find some other types of HV's to extract from LHC data!
 
 Note that such novel techniques might be used in Higgs searches if the Higgs particle is a connector to the HV \cite{HV2}.  For example, the Higgs could go to multiple $v$-hadrons, with some of those $v$-hadrons then decaying to SM muon pairs, so that one can search for the Higgs through low mass muon resonances \cite{lisanti}.
 
\begin{figure}
\begin{center}
\psfig{file=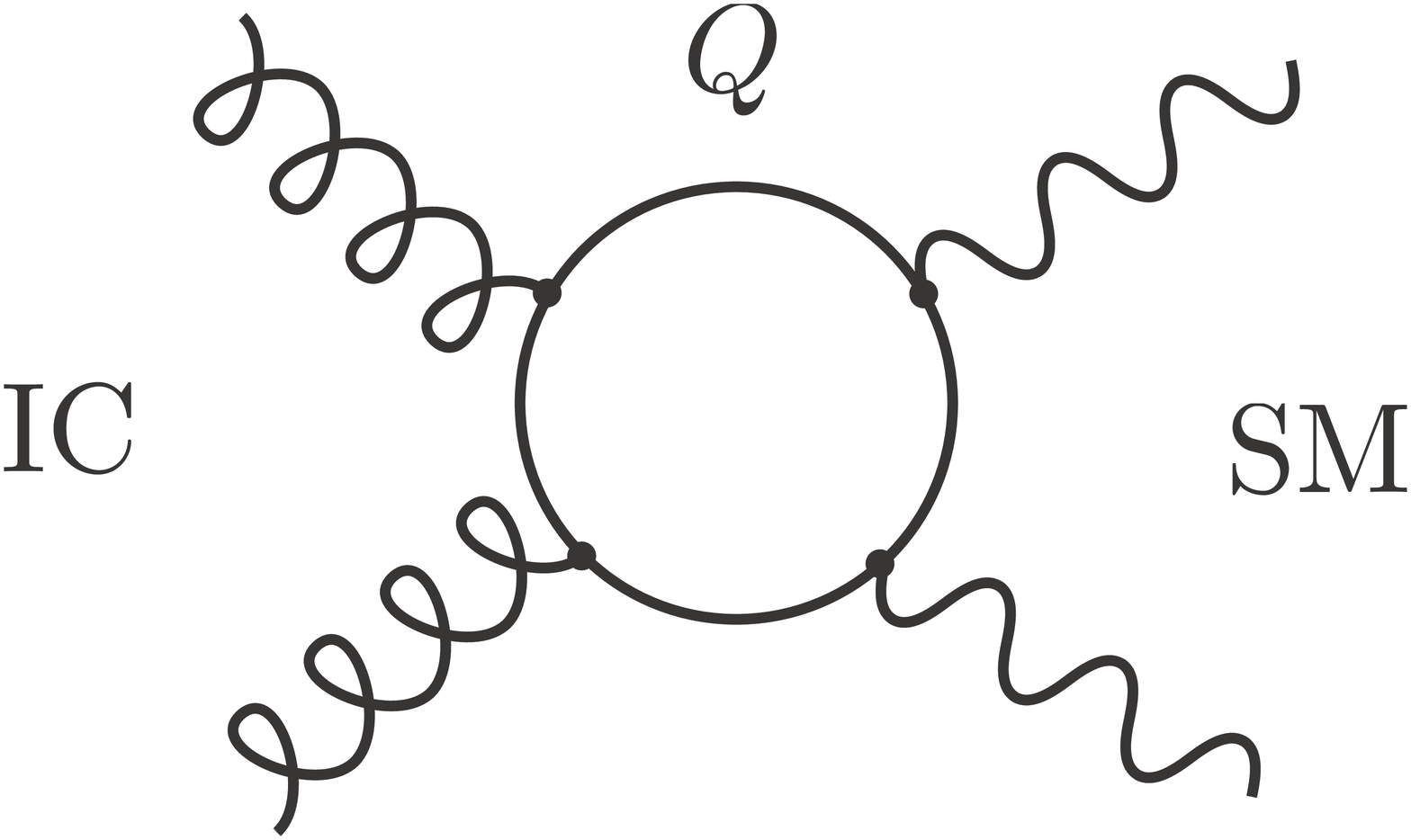,width=2.0in}
\end{center}
\caption{The quirk limit.  The hidden valley confining gauge group, called Infracolor \cite{quirks}, connects to the SM through heavy quirks which are charged under both the hidden and visible gauge groups.}
\label{QuirkMessenger}
\end{figure}

In the limit where there are {\em no} light quarks, but only a hidden sector with a low confinement scale, we come to the ``quirk'' limit \cite{quirks}.  In this limit, the connectors are typically heavy messenger quarks which are charged under both the hidden $SU(N)$ and the visible gauge groups \cite{HV,quirks}.  The process by which the quirks and hidden glueballs are produced is shown in Fig.~(\ref{QuirkMessenger}). 
 
 \section{Quirks}
 
 In the last section we discussed classes of Hidden Valleys where $m_v \ll \hat{\Lambda}$.  Now we turn to models in the opposite limit $m_v \gg \hat{\Lambda}$.  In these ``quirky'' models, the confinement scale of the hidden ``infracolor'' is anywhere between an eV and 10 GeV, while the mass scale of the heavy connector quirks is in the 100 GeV to 1 TeV range.   The ``quirks'' are charged under both infracolor and SM gauge groups.  From the diagram in Fig.~(\ref{QuirkMessenger}), the effective operators that result are
 \begin{equation}
 {\cal L}_{eff} = \frac{\hat{G}_{\mu \nu}^2 G_{\rho \sigma}^2}{M^4}, ~~~~~~\frac{\hat{G}_{\mu \nu}^2 F_{\rho \sigma}^2}{M^4},
 \label{fieldstrength}
 \end{equation}
 depending on whether the messengers are charged under SM glue or SM hypercharge.  Hats denote hidden sector field strength, and no hats indicate SM sector field strength.   The asymptotic states in the hidden sector are glueballs.
 
 The implication of these models which makes their phenomenology so unique is that there are stable, and in some cases {\em macroscopic} strings result which could be observed at the LHC.   What do we mean by stable strings and what are the implications for LHC phenomenology?  Since the mass of the quirk satisfies $m_v > \hat{\Lambda}$, the breaking of the strings is exponentially suppressed, and the length of the strings is long in comparison to $\hat{\Lambda}^{-1}$.  In order for the strings to disappear, the quirks must find each other and annihilate.  In practice, this takes many crossings.
 
 So the overall picture in the quirk limit is that quirks are pair produced, and they fly away from each other, sometimes macroscopic distances before the string pulls them back together.  They oscillate back and forth this way many times before the quirks can find each other and annihilate.  Whether the annihilation occurs in the detector and whether the string oscillations are large enough to be visible will depend on the size of the confinement scale.  Indeed we will see that the collider phenomenology will be very sensitive to the confinement scale in the hidden sector.
 
Before we move on to the LHC phenomenology, I will make a brief comment on the cosmology of these models.  First, we note that the cosmology will be safe if the reheat temperature after inflation is lower than about a GeV.  The reason for this is that the two sectors are decoupled below this temperature, for $M \sim 1 \mbox{ TeV}$.  This can be shown by comparing the rate for populating the hidden sector against the Hubble expansion $H = 1.66 g_*^{1/2} T^2/M_{pl}$.  

On the other hand, hidden glueballs can decay through the operators Eqn.~(\ref{fieldstrength}) with a rate
 \begin{equation}
 \Gamma \simeq \frac{1}{16 \pi} \frac{\hat{\Lambda}^9}{M^8}.
 \end{equation}
 This is larger than the Hubble expansion at BBN provided $\hat{\Lambda} \gtrsim 1 \mbox{ GeV}$.  This constraint can be relaxed somewhat if additional operators are added to allow for decay of hidden glueballs.

 We now discuss the various timescales relevant for the phenomenology.  Through the rest of this section in the discussion on phenomenology, we make use of Markus Luty's work and figures as presented at the Fermilab LHC Physics Center.
 
\begin{figure}
\begin{center}
\psfig{file=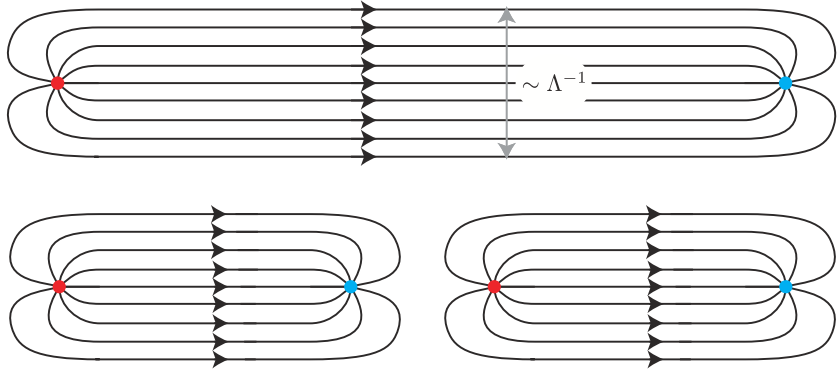,width=4.5in}
\end{center}
\caption{Strings form between quirks, which one only expects to break when the energy released in the string is larger than the quirk mass.  Since the mass of the quirk satisfies $m_v > \hat{\Lambda}$, the breaking of the string is very suppressed.}
\label{QuirkBreak}
\end{figure}

 At a collider quirks are produced in pairs with kinetic energy of order the quirk mass $m_v$.  As they are produced, they fly away from each other, and flux strings from the confinement form between the quirks, as shown in Fig.~\ref{QuirkBreak}.  The energy stored in the flux tube, $\Delta E$, is
 \begin{equation}
 \Delta E \simeq 2 m_v - \hat{\Lambda}^2 \Delta  L,
 \end{equation}
 where $\Delta L$ is the string length.  The quirks will begin to fly back together when the tension potential energy in the string becomes of order the quirk mass.  Thus we learn that the string length is
 \begin{equation}
 L \sim \frac{m_v}{\hat{\Lambda}^2} \sim 10 \mbox{ cm} \left(\frac{\hat{\Lambda}}{\mbox{keV}} .\right)^{-2}\left(\frac{m_v}{\mbox{TeV}}\right)
 \end{equation}
Since a virtual quirk-antiquirk pair has energy of order $2 m_v$, and their separation is of a size $m_v^{-1}$, we see that popping a pair out of the vacuum only lowers the string potential by an amount of order $\hat{\Lambda}^2/m_v \ll 2 m_v$, so that it is not energetically favorable to break the string.
 

\begin{figure}
\begin{center}
\psfig{file=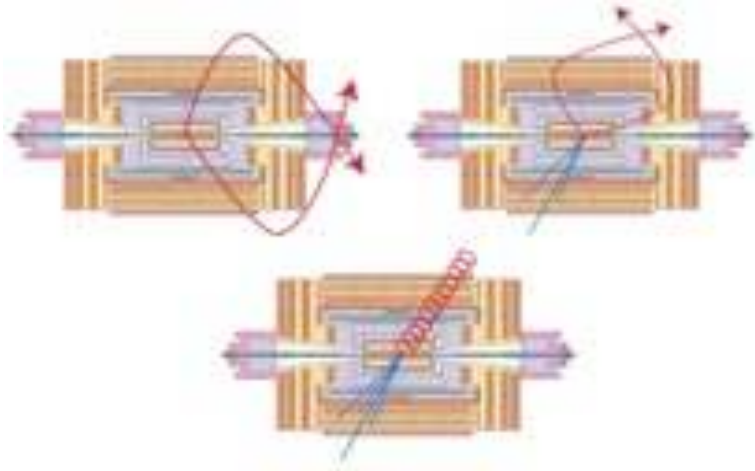,width=4.5in}
\end{center}
\caption{Macroscopic quirks.  In this case, the oscillations of the quirk strings in the detector are visible.}
\label{QuirkDetector}
\end{figure}

The phenomenology is divided by the various regimes dependent on the string length.  The first case is when
 \begin{equation}
 \mbox{mm} \lesssim L \lesssim 10 \mbox{ m} \leftrightarrow 100 \mbox{ eV} \lesssim \hat{\Lambda} \lesssim 10 \mbox{ keV}.
 \end{equation}
 In this case the quirks undergo relatively few oscillations before they exit the detector, as shown in Fig.~(\ref{QuirkDetector}).  The oscillations will be macroscopic.  Since it takes many crossings before the quirks annihilate, one only observes the tracks of the stable quirks in the detector.  If the quirks are charged, the tracks bend as they exit the detector.  
 
\begin{figure}
\begin{center}
\epsfig{file=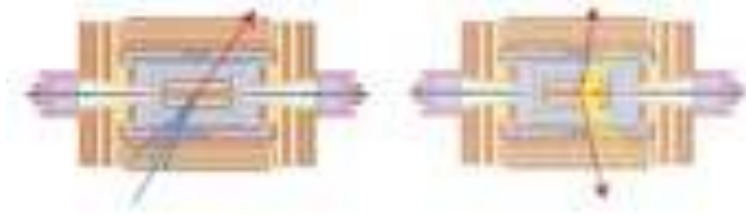,width=4.5in}
\end{center}
\caption{Mesoscopic (left) and microscopic (right) quirks.}
\label{QuirkDetector2}
\end{figure}

 The second case is mesoscopic strings,
  \begin{equation}
 \mbox{A} \lesssim L \lesssim 10 \mbox{ mm} \leftrightarrow 10 \mbox{ keV} \lesssim \hat{\Lambda} \lesssim 1 \mbox{ MeV}.
 \end{equation}
 In this case one cannot resolve the oscillations, and the quirks look like a stable charged particle.  This is shown in Fig.~(\ref{QuirkDetector2})a, recoiling against a jet.
 
The last case is microscopic strings, where microscopic is defined by comparing the string length to the quirk mass $m_v \sim 1/ \AA$.  Then in the regime
  \begin{equation}
L \lesssim \AA \leftrightarrow 1  \mbox{ MeV} \lesssim \hat{\Lambda} \lesssim 100 \mbox{ GeV},
 \end{equation}
the quirks get close enough to each other that they can annihilate and produce a mess of highly energetic photons and jets, that resemble fireballs.  This is shown in Fig.~(\ref{QuirkDetector2})b.

 \section{Unparticles}
 
 The final model we will consider which belongs to the class of low mass hidden sectors with $SU(N)$ gauge group is unparticles.  We begin here by writing down a Hidden Valley operator of the same type we have been considering up to this point, which comes from connecting a hidden sector to the SM by integrating out heavy states:
 \begin{equation}
 {\cal L}_U = \frac{1}{M_U^k}{\cal O}_{SM} {\cal O}_{BZ}.
 \label{unparticleHighDim}
 \end{equation}
 where ${\cal O}_{SM}$ is some SM operator.  The only change here as compared to a Hidden Valley is that now the hidden sector is taken to have a Banks-Zaks fixed point, represented by the operator ${\cal O}_{BZ}$ with dimension $d_{BZ}$.    Banks and Zaks constructed explicit sectors which have beta functions which run to a fixed point in the IR.  We don't really care about the details of the hidden sector ({\em i.e.} whether it exactly like the specific Banks-Zaks sector).  What is relevant is that the hidden sector is conformal in the IR, and we will call these hidden sectors BZ hidden sectors.  Recall by looking again at Table~(\ref{MS}) that BZ theories are zero $\beta$, small $\alpha N$ theories in the continuum of hidden $SU(N)$ theories.
 
 Now what Georgi did with unparticles \cite{unparticle} was to assume that the theory remains conformal in the IR, so there is no mass gap, but only a continuum of states.  If this is the case, there is no sense in which we can define particles.  To see what happens to the theory below the IR fixed point, we can match the higher dimension operator Eq.~(\ref{unparticleHighDim}) onto an unparticle operator ${\cal O}_U$ below a scale $\Lambda_U$ at which the BZ sector becomes conformal.  The effective theory below $\Lambda_U$ is then
 \begin{equation}
 \frac{\Lambda_U^{d_{BZ}-d_U}}{M_U^k} {\cal O}_{SM} {\cal O}_U,
 \end{equation}
where $d_U$ is the dimension of the unparticle operator.  

Let's see what kind of information we can get out of this sector by computing the correlator
 \begin{equation}
 \langle 0 | {\cal O}_U(x) {\cal O}_U^\dagger(0) | 0 \rangle = \int e^{-i p \dot x} | \langle 0 | {\cal O}_U | p \rangle|^2 \rho(p^2) \frac{d^4p}{(2 \pi)^4},
 \label{twopoint}
\end{equation}
where what we've done here is to insert a complete set of states, with $\rho(p^2)$ being the density of states, and evolve the operator ${\cal O}_U$ from $x$ to zero.  Noting that  ${\cal O}_U$ has dimension $d_U$, so that because of scale invariance, the matrix element $\langle 0 | {\cal O}_U(x) {\cal O}_U^\dagger(0) | 0 \rangle$ scales with dimension $2 d_U$, from which we can infer from Eq.~(\ref{twopoint}), by dimensional analysis, that 
\begin{equation}
|\langle 0 | {\cal O}_U(0) | p \rangle|^2 \rho(p^2) = A_{d_U} \theta(p^0)\theta(p^2) (p^2)^{d_U-2}.
\end{equation}
Now taking note that
\begin{equation}
(2 \pi)^4 \delta^4(p-\sum_{j=1}^n p_j) \prod_{j=1}^n \delta(p_j^2) \theta(p_j^0) \frac{d^4 p_j}{(2 \pi)^3} = A_n \theta(p^0) \theta(p^2)(p^2)^{n-2},
\end{equation}
we see that the unparticle two point correlator just gives us the phase space for $d_U$ massless particles.

Now more in line with our purpose here, we want to show the relation of unparticles with strongly coupled hidden sectors with mass gaps.  To do that we will introduce a mass gap to the unparticle, and {\em explicitly break the conformal invariance}.  To break the conformal invariance, we are going to write the continuum of unparticle states as a discrete set of states with a mass gap.  Once the mass gap is introduced, the correspondence of unparticles with strongly coupled theories with mass gaps will become more evident.  The word we will give to this process is deconstruction \cite{stephanov}.  By deconstruct, I mean to write the continuous unparticle operator as a sum of discrete states.  So let's take the Fourier transform of the two-point correlator
\begin{eqnarray}
\int d^4x e^{i p'\cdot x} \langle {\cal O}_U(x) {\cal O}_U(0) \rangle & = & \int \frac{d^4 P}{(2 \pi)^4} d^4x e^{i p'\cdot x } e^{-i P \cdot x}|\langle 0 | {\cal O}_U|P \rangle|^2 \rho(P^2) \nonumber \\ 
& = & \int d^4P \delta^4(p'-P) | \langle 0 | {\cal O}_U | P \rangle |^2 \rho(P^2)  \nonumber \\
& = & | \langle 0 | {\cal O}_U | p' \rangle |^2 \rho(p'^2)  \nonumber \\
& = & \int \frac{dM^2}{2 \pi} \rho_0(M^2) \frac{i}{p'^2-M^2+i \epsilon},
\label{unparticleQqn}
\end{eqnarray}
where
\begin{eqnarray}
 \rho_0(M^2) = 2 \pi \sum_\lambda \delta(M^2-M_\lambda^2)|\langle 0 | {\cal O}(0) | \lambda \rangle |^2.
\end{eqnarray}
 In the last step we have just inserted a delta function, $\int dM^2/(2 \pi(p'^2-M^2+i \epsilon))$.  If we define 
 \begin{equation}
 F_n^2 \equiv |\langle 0 | {\cal O}(0) | \lambda_n \rangle |^2,
 \end{equation}
 then we have
 \begin{equation}
 \rho_0(M^2) = 2 \pi \sum_n \delta(M^2 - M_n^2) F_n^2.
 \end{equation}

\begin{figure}
\begin{center}
\psfig{file=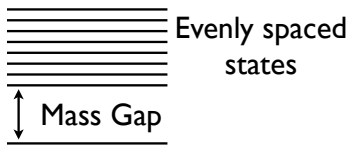,width=2.0in}
\end{center}
\caption{The mass spectrum for deconstructed unparticles.}
\label{DeconstSpec}
\end{figure}

 So finally we obtain the result that we are looking for
 \begin{equation}
 \int d^4 x e^{i P \cdot x } \langle 0 | {\cal O}(x){\cal O}(0) | 0 \rangle = \sum_n \frac{i F_n^2}{P^2 - M_n^2 + i \epsilon}.
\end{equation} 
 Now with this equation what we have shown that the unparticle correlator, when we introduce a mass gap, can be written as a sum of two point functions of discrete states of mass $M_n$.   Note that by mapping this correlator onto the unparticle correlator in the limit that the splitting between states vanishes, we can get $F_n$.  Since we know
 \begin{equation}
 \rho_O(M^2) = A_{d_U}(M^2)^{d_U-2}
 \end{equation}
 from Eq.~(\ref{unparticleQqn}), we arrive at
 \begin{equation}
 F_n^2 = \frac{A_{d_U}}{2 \pi} \Delta^2 (M_n^2)^{d_U-2},
 \end{equation}  
 where $M_n^2 = \Delta^2 n$.
 The discrete tower of states resulting from the deconstruction is shown in Fig.~(\ref{DeconstSpec}), where we have taken the freedom of decoupling the mass gap from the spectrum of evenly spaced states.  We have shown that by introducing the mass gap and writing the unparticle as a sum of discrete states, the unparticle begins to look like a Hidden Valley.
 
 But, is it exactly like a HV?  If we include only two-point correlations, and make the {\em deconstructed} unparticle a narrow tower of non-interacting resonances, they are not exactly equivalent.  The reason is that in QCD, the tower of states is self-interacting, and note with the two point correlations, one is explicitly not including any of these interactions.  To state the result more formally, a deconstructed narrow tower is only valid in the limit of large $N$.
 Large $N$ here is truly large -- if $N$ isn't in the many thousands, the interactions are dominant, and the use of the two point function gives an incomplete description of the hidden sector.
 Again referring to Table~(\ref{MS}), QCD-like Hidden Valleys are only moderate $N$, and moderate to low $\alpha N$.  
 
\begin{figure}
\begin{center}
\psfig{file=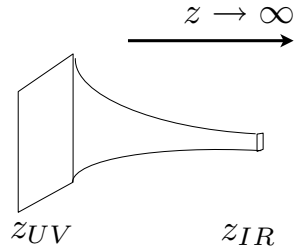,width=2.0in}
\end{center}
\caption{Extra-dimensional set-up for deconstructing unparticles.}
\label{TwoBrane}
\end{figure}

\begin{figure}
\begin{center}
\psfig{file=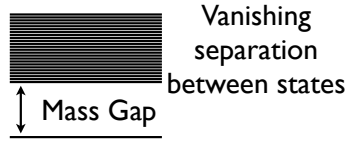,width=2.0in}
\end{center}
\caption{The unparticle spectrum we wish to reproduce with the extra dimension.}
\label{DeconstSpecConformal}
\end{figure}

 This deconstructed tower of states can be obtained explicitly from an extra dimension.  Referring to table~(1), this is valid in the extreme $\alpha N$.  In this extra-dimensional picture, the unparticles will always have interactions, since any five dimensional representation has a 5d graviton, giving rise to gravitational interactions in the bulk.  In the 4d field theory picture, this corresponds to non-negligible three point interactions.  A warped extra dimension is defined by the warped metric
 \begin{equation}
 ds^2 = ( d x^\mu d x_\mu + dz^2)/z^2.
 \end{equation}
 Note that it is conformal, as one can make the transformation 
 \begin{equation}
 z \rightarrow \alpha z,~~~~x_\mu \rightarrow \alpha x_\mu
 \end{equation}
 and the metric remains unchanged.
 For reference, we show the set-up in Fig.~(\ref{TwoBrane}).  As $z_{IR} \rightarrow \infty$, spacing between modes will vanish, and we will regain unparticle form.  Both the mass gap and spacing between the modes shown in Fig.~(\ref{DeconstSpec}) will vanish.  As the IR brane is taken to finite $z$, the conformality is broken, and the states obtain gaps between them.  The mass gap and the distance between all the states will be the same.   To reproduce an unparticle spectrum with an IR cut-off, however, with the extra dimension, we would actually need the spectrum shown in Fig.~(\ref{DeconstSpecConformal}).  Is there some way to reproduce this spectrum from the extra dimension? 
 
It's already been shown how in the literature for the case of the scalar unparticle \cite{Terning}.  Let's first see how it works for the IR brane at finite $z$.  We begin with the Lagrangian:
 \begin{equation}
 {\cal L} = \int dz \sqrt{g} \left[ g^{MN} \partial_M \Phi \partial_N \Phi - m_5^2 \Phi^2 \right]/2.
 \end{equation}
 The equation of motion which is derived from this Lagrangian is 
 \begin{equation}
 \left[\partial_z z^{-3} \partial_z + z^{-3}q^2 - z^{-5} m_5^2 \right] \Phi = 0.
 \end{equation}
 We can recast this in Schrodinger form with the definition
 \begin{equation}
 \Phi = z^{3/2} \Psi
 \end{equation}
 and we get
 \begin{equation}
 \frac{1}{z^{3/2}} \Psi'' + \frac{q^2}{z^{3/2}} \Psi - \frac{3/2 \times 5/2}{z^{7/2}} m_5^2 \Psi = 0.
 \label{XdimEOM}
 \end{equation}
 If we look at the large z limit, then we have solutions
 \begin{equation}
 \psi_n = \sin(M z + \mbox{const}).
 \end{equation}
 We can quantize these solutions with boundary conditions.  If we impose Dirichlet boundary conditions on the IR brane, for example, we must have
 \begin{equation}
 M_ n = \frac{1}{z_{IR}}(\pi n - \mbox{ const}).
 \end{equation}
 So we get a set of evenly distributed masses by deconstructing the extra dimension.
 
 Thus we've found a spectrum of unparticles related to an extra dimension, where the unparticles are deconstructed KK modes, and the spectrum is a discrete tower of weakly interacting states.   Now we would like to know whether we can recover the conformality in the UV.  The spectrum of states we would like to recover is shown in Fig.~(\ref{DeconstSpecConformal}).  This spectrum can be written very simply in the field theory as
 \begin{equation}
 \int d^4 x e^{i p \dot x} \langle 0 | {\cal O}(x) {\cal O}(0)| 0 \rangle \nonumber \\
 = \frac{A_d}{2 \pi} \int_{M^2}^{\infty}(M^2-m^2)^{d_U-2} \frac{i}{p^2-M^2+i\epsilon},
 \end{equation}
 where  we have replaced in the two-point function the term  $\sum_\lambda |\langle 0 | {\cal O} | \lambda \rangle |^2 \delta(M^2-M_\lambda^2)$ with $(M^2-m^2)^{d_U -2}$.  We must find a way to introduce {\em soft} breaking of conformal symmetry in the infrared to reintroduce the continuum.
 
 From earlier solutions, we can see that as $z_{IR} \rightarrow \infty$, the spacing between the modes vanishes.  Now, we can put in the mass gap with a profile of a field in the extra dimension.  We could have a modified equation of motion significantly modified for large $z$.  For example one can add an additional background field with profile in the extra dimension \cite{Terning}
 \begin{equation}
 H(z) = m^2 z^2
 \end{equation}
 will significantly modify the equation of motion at large $z$ in the infrared.  We then now have
 \begin{equation}
 \frac{1}{z^{3/2}}\psi'' + \frac{q^2-m^2}{z^{3/2}} \psi - \frac{3/2\times 5/2 m_5^2}{z^{7/2}} \psi = 0.
 \end{equation}  
 Such a profile could arise from an interaction $H \phi \phi$, with $H$ having scaling dimension 2.  Then we see that relative to what we had from Eqn.~(\ref{XdimEOM}), we now have
 \begin{equation}
 M_n = (q^2 - m^2)(\pi n - \mbox{ const}).
 \end{equation}
 This implies that even as $q(1/z_{IR}) \rightarrow 0$, we still have a mass gap.
 
 Thus we have shown that the features of unparticle models can be produced in a warped extra dimension, further motivating the schematic of the HV shown in Fig.~(\ref{HVthroat}).  It remains to build concrete string models of this type, with the TeV scale phenomenology of the type we have discussed in this section.
 
 \subsection{Summary: Unexpected Physics from Hidden Valley Models}
 
 We have spent  the first two lectures looking at models of hidden sectors with low confinement scales which communicate to the SM via higher dimension operators.  We've shown explicitly how each of these models, Hidden Valley, Unparticles, and Quirks, are just different faces of the same classes of models.  We've looked a little at the collider phenomenology of hidden valleys and quirks.  The phenomenology of the quirky class is particularly exotic where the confinement scale $\hat{\Lambda}$ is low, below an MeV, and stable strings can be seen to oscillate in the detector.  For HVs, the presence of light resonances in muon pairs can be a particularly striking signal.  Displaced vertices are common, as well as missing energy from stable $v$-hadrons.  In the unparticle limit, where the mass splittings between states is taken to zero, the states in the hidden sector are stable, so that more conventional missing energy searches should suffice.  What should be clear, however, is that a relatively small class of these models has already been explored in detail, leaving much room for exploration.

 \section{Models of Hidden Sector Dark Matter}
 
\begin{figure}
\begin{center}
\psfig{file=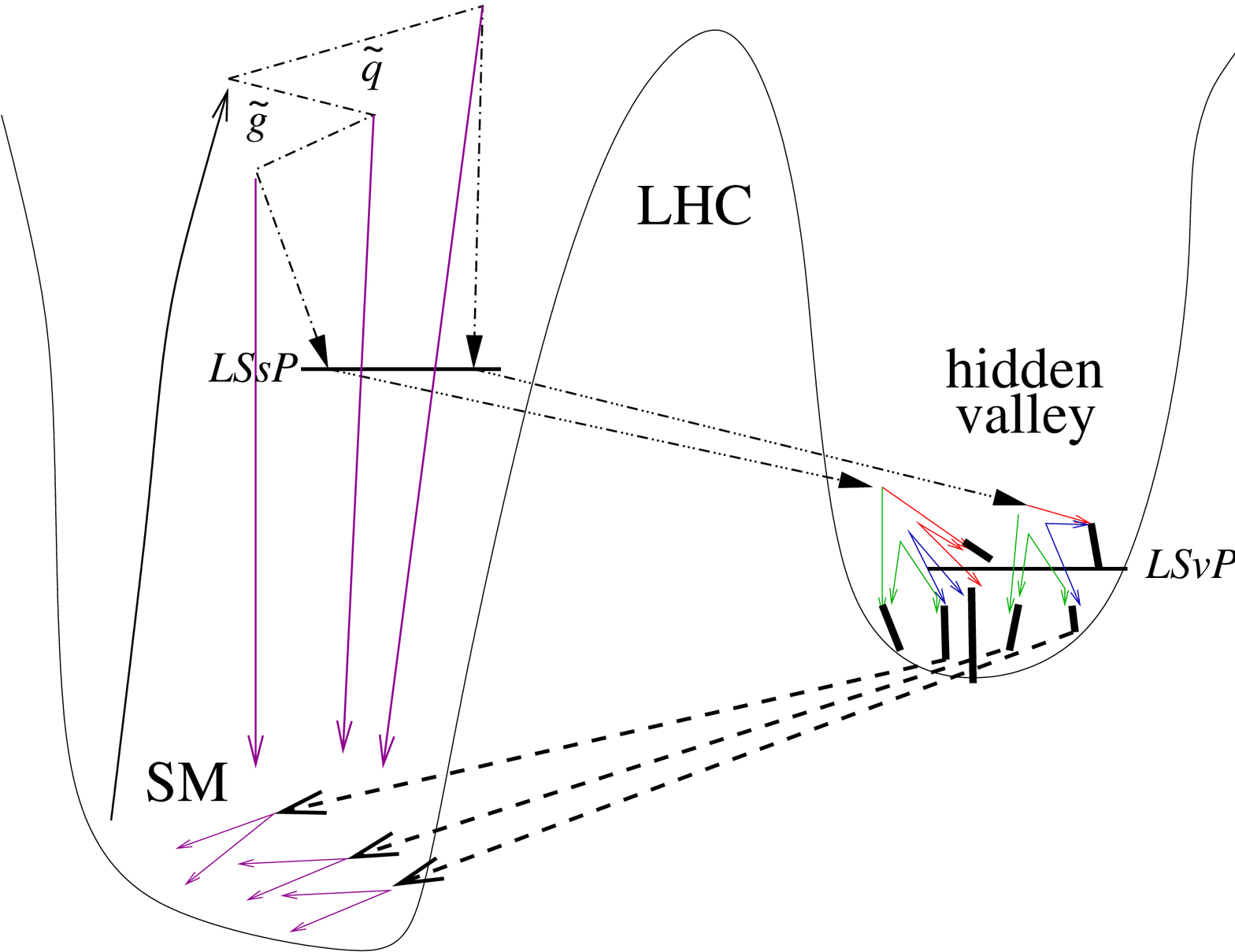,width=3.5in}
\end{center}
\caption{The effects of supersymmetry on a hidden valley.}
\label{HVSUSY}
\end{figure}

Hidden Valleys have potentially important implications for Dark Matter.  First, as suggested by Fig.~(1), the dark sector may have complex dynamics -- it may not contain a single stable weakly interacting particle.  Within the context of supersymmetry, the presence of the HV causes the lightest supersymmetric particle to be unstable to decay to hidden sector particles \cite{SUSYHV} .  That is to say, that the lightest supersymmetric particle no longer resides in the visible MSSM sector, but instead in the low mass hidden sector, as shown in Fig.~(\ref{HVSUSY}).  If this is the case, then dark matter dynamics within the context of supersymmetry can be changed dramatically.  In this section we significantly broaden and extend this notion that dark matter dynamics can be modified significantly.  
 
So let's begin our discussion about HV dark matter by examining our prejudices about dark matter to see whether they are really very well founded.  What we have been taught more or less believe about dark matter is that it is
\begin{itemize}
\item single -- made up predominantly of one component
\item stable
\item weakly interacting
\item neutral
\item weak scale
\end{itemize}
particle.
This is something of a ``spherical cow'' approximation of dark matter.  Now on what basis are these notions based?  They are not totally unfounded, so let's go through the reasons.
\begin{itemize}
\item The dark matter is a single state.  In most models this has been true for two reasons.  First, there is usually one new symmetry, such as $R$-parity, so only one new state.  However, if we are willing to widen our field of view to hidden sectors with complex dynamics, there may be additional symmetries, such as dark lepton and dark baryon number to keep additional particles stable.  The other reason it has been argued that the dark matter is predominantly one component is that thermal freeze-out calculations, which we will review below, would seem to indicate that having more than one component of dark matter with the same density would be rather tuned.  We will show explicitly that this need not be the case {\em a priori}. 
\item The dark matter is stable.  I don't think this is such a bad assumption.  Dark matter exists in the universe today, so it's stable or at least long lived.
\item The dark matter is weakly interacting.  If it was much stronger than weakly interacting, we would have already seen it in direct detection experiments.  If it's much more weakly interacting than weak, we're going to have a very difficult time detecting it in any direct fashion at all.  This is possible, but we're going to assume that  we have some hope of seeing it directly, and that the weak scale is a well-motivated place to look for dark matter.  On the other hand, the dark matter need not be so weakly interacting with itself.  That is to say, the dark matter could have dark forces which give rise to significant effects in the dark sector, and indeed this is quite likely in HV models, since the confining gauge group is itself a dark force.
\item The dark matter is electrically neutral.  There are strong constraints on the charge of the dark matter.  But if we are willing to widen our field of view to models of dark matter with strong dynamics, the constituents of the  neutral dark bound state might in fact be charged. As we will see, this can be quite natural in Quirky dark matter models.  The direct detection signals are unique in that case.
\item The dark matter is a weak scale particle.  Much of the motivation for focusing on weak scale dark matter has to do with the fact that the thermal freeze-out calculation, which, again, we will review below, suggests that the weak scale gives rise to dark matter with the observed relic density.  However, in HV models, one might expect dark matter components which are much lighter.  Does this ruin the coincidence of the thermal freeze-out calculation?  Not necessarily, as we will see in two separate cases.  
\end{itemize}

\begin{figure}
\begin{center}
\psfig{file=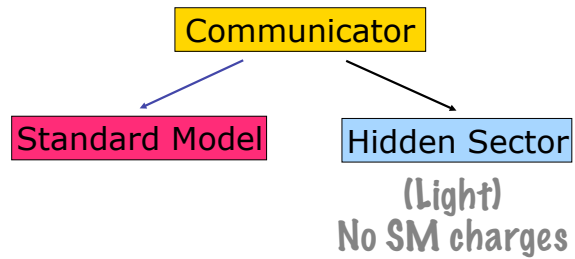,width=3.0in}
\end{center}
\caption{The class of models we are considering for dark matter.  The hidden sector may or may not have confining dynamics.}
\label{HVBlock}
\end{figure}

Many of the statements here are rather vague, as we have many possible realizations for HV models, including examples which have yet to be built (perhaps by you!).  So what we are going to do now is to go through a few examples which are well motivated for solving particular theoretical or observation problems.  Some of these models have explicitly strong dynamics in the hidden sector, while other models simply contain low mass hidden sectors which communicate to the standard model through states which couple to both sectors, as shown in the schematic of Fig.~(\ref{HVBlock}).  In all cases, the dark sectors have non-trivial dynamics with multiple states and dark forces, whether Abelian or non-Abelian.  

The choice here is a personal one, but I hope you will bear with me since this is the second to last lecture on the last day of TASI.

\subsection{Light Abelian Hidden Sectors}

\begin{figure}
\begin{center}
\psfig{file=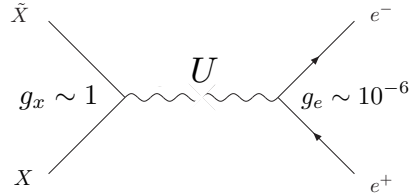,width=2.5in}
\end{center}
\caption{The matter content of the MeV dark matter model.  The dark matter field is $X$ (which may be a scalar or a fermion), and the gauged MeV mediator is $U$.}
\label{mev}
\end{figure}

Models of MeV dark matter \cite{fayet} fit into the HV paradigm \cite{MeV}, in the sense that they contain low mass hidden sectors with dark forces which couple to the SM model weakly.  Phenomenologically, the MeV dark matter model was postulated some time ago \cite{fayet} to explain the observed excess of 511 keV radiation toward the galactic center observed by SPI/Integral.  Fayet built a model where the signal could be produced by dark matter in the galactic center annihilating to $e^+ e^-$ pairs, which in term annihilate to 511 keV radiation.  The matter and gauge content and couplings are shown in Fig.~(\ref{mev}).   From a model building point of view this model looks somewhat, shall we say, contrived: the model contains MeV dark matter, an MeV gauged $U(1)$ mediator, ${\cal O}(1)$ coupling of the mediator to the dark matter, and ${\cal O}(10^{-6})$ coupling of the mediator to electrons.  What could generate such a dark sector?

\begin{figure}
\begin{center}
\psfig{file=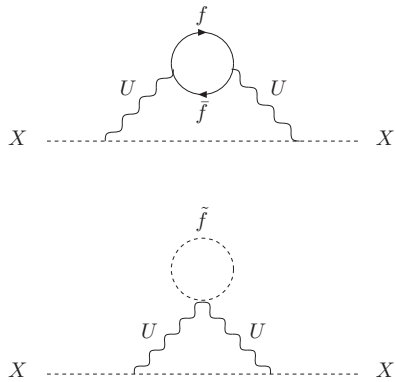,width=3.0in}
\end{center}
\caption{Communication of SUSY breaking through two loop graphs.  The dark sector scalars $X$ receive SUSY breaking contributions by communicating through the light mediator $U$ to MSSM fermion $f$ and its scalar super-partner $\tilde{f}$ \cite{MeV}.}
\label{twoloop}
\end{figure}

\begin{figure}
\begin{center}
\psfig{file=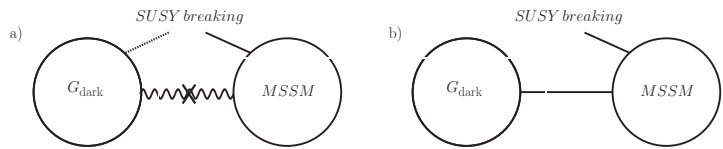,width=4.0in}
\end{center}
\caption{A schematic of hidden sector models which communicate through kinetic mixing \cite{arkani}.  This set-up has been considered in multiple contexts \cite{MeV,feng,pospelov,arkani}.}
\label{Communication}
\end{figure}

The interest in the model as a HV started with a simple observation, namely that
\begin{equation}
\mbox{MeV} \sim 10^{-6} \mbox{ TeV}.
\end{equation}
That is, we wish to connect TeV scale supersymmetry breaking to the MeV hidden sector through the small coupling between the mediator the MSSM.  Normally one expects that hidden sector soft SUSY breaking masses will be around the TeV scale along will all the other superpartner masses.  This is generally true if {\em gravity} mediation generates the soft SUSY masses, since, in the absence of sequestering, gravity couples equally to all states, hidden or visible.  However, if gauge mediation generates the soft SUSY masses, the hidden sector can be shielded from MSSM gauge mediated SUSY breaking masses by small couplings to MSSM states.  Then one expects the SUSY breaking masses could be much smaller.  Take for example the two loop diagram in Fig.~(\ref{twoloop}), where we imagine that MSSM states $f$ and $\tilde{f}$ are running in the loop.  This diagram generates soft masses for states in the hidden sector which are of the size
\begin{equation}
m_{x}^2 = q_x^2 q_I^2 \left( \frac{g_x}{g_e}\right)^2 m_{E^c}^2,
\label{1loop}
\end{equation}
where $g_{x,e}$ are the gauge couplings of the dark sector $x$ and electron $e$ to the gauged mediator, $q_x$ and $q_l$ the charges, and $m_{E^c}$ is the SUSY breaking mass of the right-handed selectron.  Since $g_e \sim 10^{-6}$ and $g_x \sim 1$ in the MeV dark matter model, we can see that the MeV scale naturally comes about \cite{MeV,arkani,wang,poland}.  Now why would the dark sector and electrons have such different couplings to the mediator?  This can happen through kinetic mixing between hypercharge and the new hidden $U(1)$ \cite{MeV,arkani}, so that
\begin{equation}
g_e = g_Y \epsilon,
\label{epsilon}
\end{equation}
where $g_Y$ is the hypercharge gauge coupling, and $\epsilon$ is the coefficient of the kinetic mixing term
\begin{equation}
{\cal L}_{kin} = \epsilon F_{\mu \nu} \tilde{F}^{\mu \nu}.
\end{equation}
A schematic drawing of this type of communication between sectors is shown in Fig.~(\ref{Communication}).

Now we have seen how one can build natural particle physics models with MeV mass scalars, gauge fields, and small couplings to the SM.  However, we have been taught that one of the most compelling reasons for considering weak scale dark matter is that we naturally obtain the right relic abundance, and this phenomenon is called the ``WIMP miracle.''  So what about the MeV dark matter model -- is there any miracle that we obtain the right relic abundance for MeV dark matter?  First, let's review the thermal freeze-out arguments that constitute the WIMP miracle.   The dark matter number density freezes out at a temperature $T_{fo}$ when the annihilation cross-section becomes of order the Hubble expansion:
\begin{equation}
n_{DM} \sigma_{ann} v \lesssim H(T_{fo}) = 1.66 g_*^{1/2} T_{fo}^2/M_{pl}.
\end{equation}
Now we use the fact that (up to some constants) the dark matter energy density at freeze-out $\rho_{DM}(T_{fo})$ is
\begin{equation}
\rho_{DM} \sim T_{fo}^4 \left( \frac{m_{DM}}{T_{fo}}\right)^{5/2} e^{-m_{DM}/T_{fo}}.
\end{equation}
Now if we compare this against the observed dark matter to photon energy density ratio
\begin{equation}
\eta_\gamma \equiv \frac{\rho_{DM}^0}{T_\gamma^4} = \frac{T_{fo}}{T_\gamma}\left(\frac{m_{DM}}{T_{fo}}\right)^{5/2} e^{-m_{DM}/T_{fo}},
\end{equation}
where $\rho_{DM}^0$ is the observed dark matter density today and $T_\gamma$ the CMB photon energy today,
we can solve for $m_{DM}/T_{fo} \approx 1/25$ (which is logarithmically sensitive to the overall scale $T_{fo}$).  Using this ratio, we can then solve for the dark matter annihilation cross-section needed to obtain the observed relic density:
\begin{eqnarray}
\sigma_{ann} v&  \sim & \frac{m_{DM}}{T_{fo}}\frac{T_\gamma^3}{M_{pl}} \frac{1}{\rho_{DM}^0} \nonumber \\
& \simeq & 3 \times 10^{-26} \mbox{cm}^3/\mbox{s}.
\end{eqnarray}
If one does the dimensional analysis on this cross-section one finds it is 
\begin{equation}
\sigma_{ann} v \simeq \frac{1}{\mbox{TeV}^2},
\end{equation}
from which many have concluded the weak scale dark matter is very well motivated, since typically annihilation cross-sections scale as
\begin{equation}
\sigma_{ann} v \simeq \frac{g^4}{m_{DM}^2},
\label{relic}
\end{equation}
where $g$ is the coupling of the dark matter to the mediator of the annihilation.  It is typically ${\cal O}(1)$.

So do we destroy this relation for dark matter in a hidden valley well below a TeV? Not necessarily, and not for MeV dark matter.  In the case of hidden sectors connected to the visible sector via gauge mediation, this is particularly natural.  From Eqs.~(\ref{1loop}),~(\ref{epsilon}), we see that  for the MeV model the annihilation diagram in Fig.~(\ref{mev}) scales so that $g^4 = g_x^2 g_e^2 \sim (10^{-6})^2$.  Since $m_{DM} \sim 10^{-6} \mbox{ TeV}$, from Eq.~(\ref{relic}) we see that the correct relic abundance is obtained.  Thus in abelian hidden sectors weakly coupled to the standard model through light gauged mediators, the WIMP miracle can be preserved, and one sees why the correct relic density is still obtained for MeV dark matter, even though it is not a weak scale dark matter candidate.  

In general, in hidden sectors where SUSY breaking is communicated to it through gauge mediation, the relation $m_{DM} \sim g^2 m_{SUSY}$ is obtained.  As Feng and Kumar observed, it is a ``WIMPless'' miracle \cite{feng}.  With multiple stable particles one can now see why it might be natural for multiple stable particles to have comparable relic densities, if they satisfy Eq.~(\ref{relic}).

Now these hidden sector models have become relevant to the PAMELA and FGST cosmic ray excesses because light dark forces can give rise to boosted annihilation cross-sections when the dark matter becomes non-relativistic \cite{Nojiri,MinimalDM,TODM}.  Though I do not have the time to go into these models in detail here, suffice it to say that these models require light mediators, 1 GeV or lighter, if the dark matter itself is weak scale.  Given what we've learned now about how naturally light gauged mediators can arise, this implies that the natural size of this coupling is $\epsilon \simeq 10^{-3}$ \cite{arkani,wang}.

The exploration of these classes of models has subsequently given rise to fruitful discussions on detection of light dark forces in high luminosity fixed target and $e^+ e^-$ colliders \cite{gevexperiments}.

Lastly, since the models we have discussed are supersymmetric, we note that the collider phenomenology of Fig.~(\ref{HVSUSY}) results in these models.  The MSSM LSP is unstable to decay to the LSP in the SUSY sector.  The lifetime of the LSP is model dependent, but because the coupling to the hidden sector is so small, the lifetimes of the MSSM LSP tend to be long, and displaced vertices can appear in the detector from the LSP decays.

\subsection{Solutions to the Baryon Dark Matter Coincidence}

Now let's consider another type of model where the dark matter resides in a hidden sector and has a mass well below a TeV, though its density is not set by thermal freeze-out.

Phenomenologically we know that $\rho_{DM}/\rho_b \approx 5$, though within the standard paradigm of thermal freeze-out there is no explanation for this ratio, since the dark matter and baryon densities are set through very different mechanisms, the former by thermal freeze-out (determined by the dark matter mass and its coupling to standard model states), and the latter through CP violating phases and out-of-equilibrium dynamics.  Now one could imagine writing down some mechanism where the two number densities are related to each other (we explain how in a minute)
\begin{equation}
n_x - n_{\bar{x}} \approx n_b - n_{\bar{b}},
\end{equation}
where now we are generating an asymmetry between the dark matter number density $n_x$ and the anti-dark matter number density $n_{\bar{x}}$.  If this is the case, then the observed energy densities of dark matter implies
\begin{equation}
\frac{m_x}{m_p} \approx 5.
\end{equation}
For concrete models, the dark matter is typically not precisely 5 GeV, since charges enter into the relation between the $x$ asymmetry and the baryon asymmetry, but usually they are quite close to each other.

This fact by itself is enough to put this low mass dark matter into the class of HV which communicate through a heavy mediator, as shown in Fig.~(\ref{SUSYValley})).  There may or may not be strong dynamics in the hidden sector.

We write down operators of the form Eq.~(\ref{unparticleHighDim}).
The idea behind these models is to write an effective field theory which describes the interactions between the hidden sector and visible sector (integrating out the fields residing at the ``pass'' in Fig.~(1), which transfers a Standard Model baryon or lepton asymmetry to the dark sector.  The dark matter in these models must be sterile, so this limits the number of operators which can be constructed to accomplish this purpose.  In particular, in the context of supersymmetry, the lowest dimension operators carrying lepton or baryon number which are sterile are
\be
W = {\cal O}_d udd \\ \nonumber
W = {\cal O}_d LH,
\ee
where ${\cal O}_d$ is an operator for dark sector fields.
If these operators are connected to the hidden sector containing the dark field $\bar{X}$ to transfer an asymmetry, we have \cite{ADM}
\be
\label{ops}
W = \frac{\bar{X}^2 udd}{M^2} \\ \nonumber
W = \frac{\bar{X}^2 LH}{M}.
\ee
The second operator, for example, enforces $2(n_X - n_{\bar{X}}) = n_{\bar{\ell}} - n_{\ell}$, and a detailed calculation relating the lepton asymmetry to the baryon asymmetry (through sphalerons) consequently shows that this model predicts $m_X \simeq 8 \mbox{ GeV}$.  Note that we added $\bar{X}^2$ and not $X$, since the additional $Z_2$ symmetry ensures DM stability.  In some other cases, $R$-parity may be utilized instead to stabilize the dark matter

Now once the Standard Model baryon or lepton asymmetry has been transferred to the dark sector, the symmetric part of the dark matter (which is much larger than the asymmetric part, $n_X + n_{\bar{X}} \gg n_X - n_{\bar{X}}$) must annihilate, leaving only the asymmetric part.  There are a variety of mechanisms to do this, but the difficulty here is having a mechanism which is efficient enough to annihilate away the whole of the symmetric part through $X \bar{X} \rightarrow SM$.  Such a process, through a dimension six operator has a cross-section
\be
\sigma v = \frac{1}{16 \pi} \frac{m_X^2}{M'^4}.
\ee
This cross-section must be bigger than approximately 1 pb in order to reduce the dark matter density to its asymmetric component, implying $M' \lesssim 100 \mbox{ GeV}$, a rather severe constraint for any new electroweak state coupling to Standard Model states.

Here confinement in the hidden sector can be a useful tool.  If the dark matter consists of symmetric and asymmetric bound states of elementary dark sector fermions, the symmetric states may decay through the same dimension six operators, while the asymmetric states would remain stable.  For example, suppose in the operator Eq.~(\ref{ops}), we replaced the operator $\bar{X}^2$ with $\bar{v}_1 v_2$, and supposing these $v_1$ and $v_2$ constituents are charged under a hidden sector confining gauge group, such that bound states $\bar{v}_1 v_2$, $\bar{v}_2 v_1$ and $\bar{v}_1 v_1 + \bar{v}_2 v_2$ are the relevant degrees of freedom at low energies.  When Eq.~(\ref{ops}) freezes out, the asymmetric $\bar{v}_1 v_2$ states remain stable, while the symmetric $\bar{v}_1 v_1 + \bar{v}_2 v_2$ states decay rapidly through less suppressed operators (that is, we take $M' \ll M$).  In the next section we describe a related class of confinement models where the constituents of the dark matter bound states carry electroweak charges.  In these models sphalerons rather than higher dimension operators such as Eq.~(\ref{ops}) to transfer the asymmetry.

\subsection{Composite Dark Matter}

Lastly we consider an honest HV with honest confinement in the dark sector \cite{QDM}.  In this case, we are imagining that the constituents are electroweak charged, but that the dark matter is a neutral bound state of electroweak charged quirks \cite{QDM}.  The constituents must have electroweak scale masses, but the confinement scale of the dark gauge group can be much below the weak scale.  That is, the dark matter in the model is a bound state of quirks.  The electroweak charges of the constituents are chosen so that we can transfer asymmetry between sectors using the electroweak sphalerons.  The dark matter and baryon asymmetries really just get mixed up with each other through the electroweak sphalerons, so that the baryon and dark matter asymmetries are related.  The charges of the constituents are shown in Table~(\ref{table:model}).  

\begin{table}
\begin{center}
\begin{tabular}{c|cccc}
        & $SU(2)_{Q}$ & $SU(2)_L$ & $U(1)_Y$ & $U(1)_{QB}$ \\ \hline
$\xi_Q = \begin{pmatrix} \xi_U, \xi_D\end{pmatrix}$     &  
        $\mathbf{2}$      &   $\mathbf{2}$     &   $0$  &  $+1/2$  \\
$\xi_{\bar{U}}$ &     $\mathbf{2}$      &    -      &   $-1/2$ & $-1/2$ \\
$\xi_{\bar{D}}$ &     $\mathbf{2}$      &    -      &   $+1/2$ & $-1/2$ 
\end{tabular}
\end{center}
\caption{Particle content and charges under the gauge and global symmetries.}
\label{table:model}
\end{table}

In particular, the sphalerons will violate some linear combination of $B$, $L$ and dark baryon number, $DB$.  Thus an asymmetry in $B$ and $L$ (produced from some leptogenesis or baryogenesis mechanism) will be converted to an asymmetry in $DB$.  The $DB$ asymmetry then sets the dark matter relic density.  Since the dark matter mass is around the mass of the weak scale quirk constituents, there must be a Boltzmann suppression in $DB$ to achieve the observed relation $\Omega_{DM} \simeq 5 \Omega_b$.  This can be naturally achieved when the sphalerons decouple just below the dark matter mass:
\be
\Omega_{DM} \sim \frac{m_{DM}}{m_p} e^{-m_{DM}/T_{sph}} \Omega_b,
\ee
where $T_{sph}$ is the sphaleron decoupling temperature.  

Now we come back to the neutral dark matter question.  The dark matter bound state does not carry electric charge, but since its constituents do, the dark matter itself still couples to the photon.  So what about photon absorption on these states, and large scattering cross-sections at direct detection experiments?  It turns out (see \cite{QDM} for details) that the coupling of the photon to the neutral dark state depends on the mass splitting between the constituent quirks, so that to evade constraints from an experiment like CDMS, the mass splitting between the constituent quirks must be smaller than about one part in $10^3$.  This is just telling you that the charge radius vanishes as the wavefunctions of the constituents in the bound state become identical, {\em i.e.} charge cannot be resolved by the photon.  There are other effects to worry about, such as the fact that an external electric or magnetic field (i.e. a photon) can polarize the bound state anyway and give rise to scattering through the photon again.  How easy it is for the photon to do this depends of course on how tightly the quirks are bound together by the dark color force ({\em i.e.} what the Bohr radius for the dark color force is). The other interesting fact about these models is that they can absorb photons on galactic scales, though it requires a rather large density of dark matter.  This latter fact is something that remains to be investigated in more detail.


I have only sketched the details of this model, but you can see that by looking at composite dark matter you can find some pretty interesting phenomenology.  People have also been looking at composite dark matter to give rise to small mass splittings in the bound states.  They want to do this so that they can generate inelastic scattering of dark matter off nuclei, where the dark matter only scatters on nuclei when the interaction has enough energy to kick the dark matter into the excited state.  Tucker-Smith and Weiner have tried to use such mass splittings to reconcile the positive result of DAMA with the null results of other experiments.  They postulate that the heavy iodine gives the recoiling dark matter enough of a kick to boost it into an excited state, while interactions with lighter nuclei such as Germanium do not result in enough momentum transfer to knock the dark matter into the excited state, giving rise to no signal in a Germanium experiment such as CDMS.  I know of no concrete model of composite inelastic dark matter in the literature (abelian models exist), though {\em phenomenological} models of composite inelastic have been derived.  Building such a model can be your homework problem.

\subsection{Summary: Hidden Valley Dark Matter}

\begin{figure}
\begin{center}
\includegraphics[width=8cm]{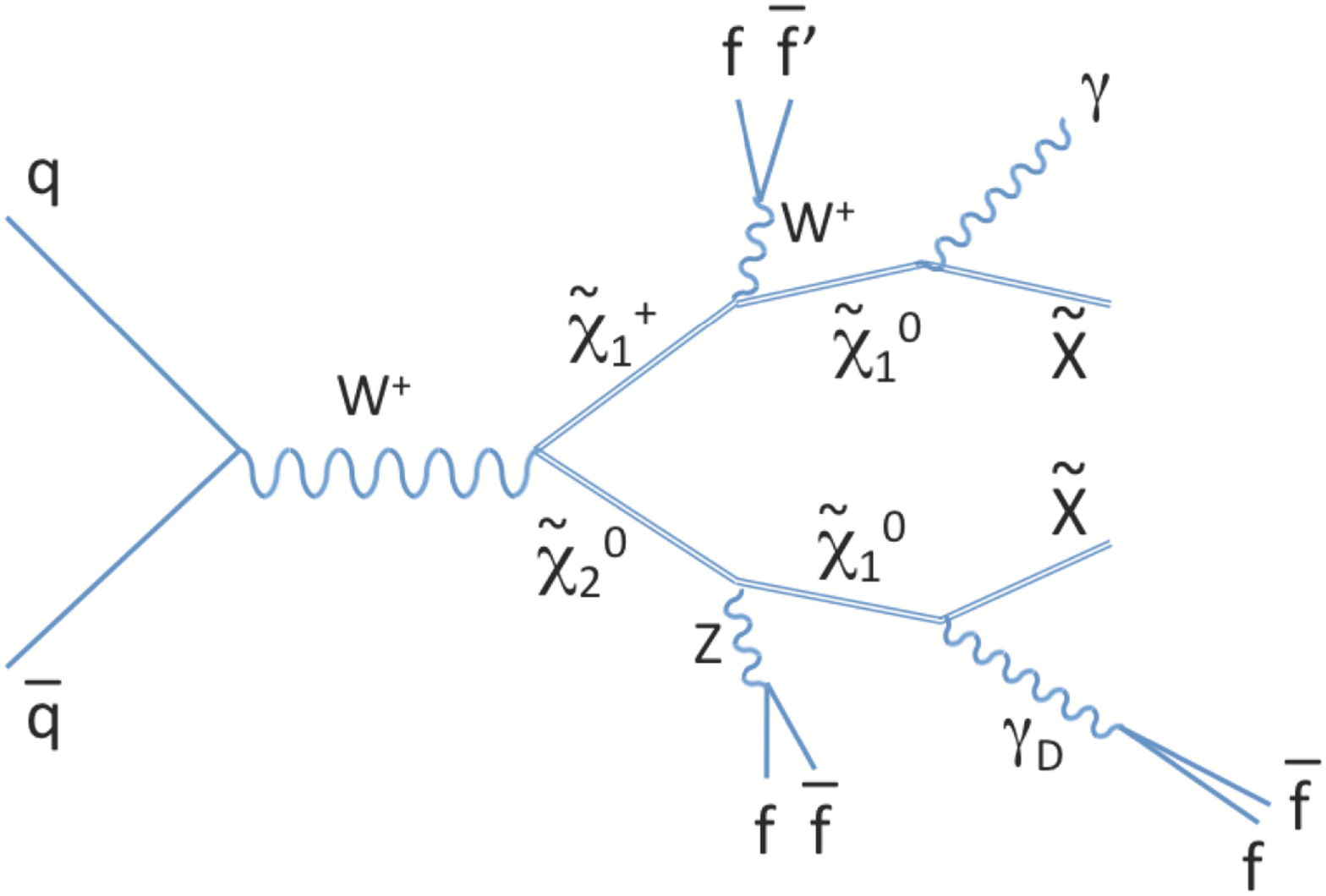}
\end{center}
\caption{\label{fig:diagram1} One of the diagrams giving rise to the events with a photon, dark photon ($\gamma_D$), 
and large missing energy due to escaping darkinos ($\tilde{X}$) at the Fermilab Tevatron Collider \cite{D0}. }
\end{figure}

As you can see, there are many possibilities with the structure Eq.~(\ref{unparticleHighDim}).  We have looked at a  few possibilities with Abelian and non-Abelian hidden sectors.  Perhaps one of the most interesting consequences of this direction for model building is that the experiments are now beginning to direct more effort towards detecting dark matter candidates from hidden valleys.  The experiments now are beginning to look for low mass dark photons, for example, which mix with the visible photon, and decay to muon pairs.  As we illustrated in Fig.~(\ref{HVpheno})b, such searches for low mass resonances in muon pairs can be very efficient in reducing or eliminating SM backgrounds.  Taken from a D0 search for Hidden Valleys \cite{D0}, we show in Fig.~(\ref{fig:diagram1}) a type of event which may produce dark matter.  

 \section{Conclusions}
 
 Hidden Valleys are no longer totally unexpected at the LHC, so the title of the lecture is something of a misnomer.  However, in comparison to the mostly studied models of new physics at the LHC, which have focused on resonances at the electroweak scale, they remain still relatively unknown, and deserve further attention as the LHC commissioning goes forward.  The unexpected physics may mean searching for light resonances which decay to muon pairs in an otherwise high center of mass energy event, or searching for SUSY with a light hidden sector where the MSSM LSP decays to the hidden sector particles with a displaced vertex.  In either case, new search techniques will be necessary for uncovering the new physics.  However, it is also true that the parameter space of hidden sector models has not been fully explored, and many models can be built, also by you.  It is important in this data driven era, however, to remain focused on phenomenology and signals which may be searched for at the LHC, and in direct and indirect detection of dark matter experiments.  We may find that we will shortly be uncovering new physics of a nature we had not previously anticipated.

\bibliographystyle{ws-procs9x6}
\bibliography{thebibliography}

\begin{thebibliography}{9}

\bibitem{HV}
  M.~J.~Strassler and K.~M.~Zurek,
  Phys.\ Lett.\  B {\bf 651}, 374 (2007)
  [arXiv:hep-ph/0604261].
  
\bibitem{quirks}
  J.~Kang and M.~A.~Luty,
  JHEP {\bf 0911}, 065 (2009)
  [arXiv:0805.4642 [hep-ph]].
  
\bibitem{unparticle}
  H.~Georgi,
  Phys.\ Rev.\ Lett.\  {\bf 98}, 221601 (2007)
  [arXiv:hep-ph/0703260].
    H.~Georgi,
  Phys.\ Lett.\  B {\bf 650}, 275 (2007)
  [arXiv:0704.2457 [hep-ph]].
  
  \bibitem{THanTASI}
    T.~Han,
  arXiv:hep-ph/0508097.
  
  \bibitem{refMS}
  This table is based on a talk by Matt Strassler.
  
  \bibitem{HVpheno}
  T.~Han, Z.~Si, K.~M.~Zurek and M.~J.~Strassler,
  JHEP {\bf 0807}, 008 (2008)
  [arXiv:0712.2041 [hep-ph]].
  
\bibitem{HVheavy}
  M.~J.~Strassler,
  arXiv:0806.2385 [hep-ph].
  
\bibitem{HV2}
  M.~J.~Strassler and K.~M.~Zurek,
  Phys.\ Lett.\  B {\bf 661}, 263 (2008)
  [arXiv:hep-ph/0605193].
  
  \bibitem{lisanti}
    M.~Lisanti and J.~G.~Wacker,
  Phys.\ Rev.\  D {\bf 79}, 115006 (2009)
  [arXiv:0903.1377 [hep-ph]].
  
\bibitem{SUSYHV}
  M.~J.~Strassler,
  arXiv:hep-ph/0607160.
  

   \bibitem{stephanov}
  M.~A.~Stephanov,
  Phys.\ Rev.\  D {\bf 76}, 035008 (2007)
  [arXiv:0705.3049 [hep-ph]].
  
\bibitem{Terning}
  G.~Cacciapaglia, G.~Marandella and J.~Terning,
  JHEP {\bf 0902}, 049 (2009)
  [arXiv:0804.0424 [hep-ph]].
  
 \bibitem{fayet}
   C.~Boehm and P.~Fayet,
  Nucl.\ Phys.\  B {\bf 683}, 219 (2004)
  [arXiv:hep-ph/0305261].
   P.~Fayet,
  Phys.\ Rev.\  D {\bf 70}, 023514 (2004)
  [arXiv:hep-ph/0403226].
  
\bibitem{MeV}
  D.~Hooper and K.~M.~Zurek,
  Phys.\ Rev.\  D {\bf 77}, 087302 (2008)
  [arXiv:0801.3686 [hep-ph]].
  
\bibitem{arkani}
  N.~Arkani-Hamed and N.~Weiner,
  JHEP {\bf 0812}, 104 (2008)
  [arXiv:0810.0714 [hep-ph]].
 
\bibitem{wang}
  C.~Cheung, J.~T.~Ruderman, L.~T.~Wang and I.~Yavin,
  Phys.\ Rev.\  D {\bf 80}, 035008 (2009)
  [arXiv:0902.3246 [hep-ph]].
  
\bibitem{poland}
  D.~E.~Morrissey, D.~Poland and K.~M.~Zurek,
  JHEP {\bf 0907}, 050 (2009)
  [arXiv:0904.2567 [hep-ph]].
  
\bibitem{feng}
  J.~L.~Feng, J.~Kumar and L.~E.~Strigari,
  Phys.\ Lett.\  B {\bf 670}, 37 (2008)
  [arXiv:0806.3746 [hep-ph]].
  
\bibitem{pospelov}
  M.~Pospelov, A.~Ritz and M.~B.~Voloshin,
  Phys.\ Lett.\  B {\bf 662}, 53 (2008)
  [arXiv:0711.4866 [hep-ph]].
  
\bibitem{Nojiri}
  J.~Hisano, S.~Matsumoto, M.~M.~Nojiri and O.~Saito,
  Phys.\ Rev.\  D {\bf 71}, 063528 (2005)
  [arXiv:hep-ph/0412403].
  
\bibitem{MinimalDM}
  M.~Cirelli, M.~Kadastik, M.~Raidal and A.~Strumia,
  Nucl.\ Phys.\  B {\bf 813}, 1 (2009)
  [arXiv:0809.2409 [hep-ph]].
  
\bibitem{TODM}
  N.~Arkani-Hamed, D.~P.~Finkbeiner, T.~R.~Slatyer and N.~Weiner,
  Phys.\ Rev.\  D {\bf 79}, 015014 (2009)
  [arXiv:0810.0713 [hep-ph]].
  
  \bibitem{gevexperiments}
    J.~D.~Bjorken, R.~Essig, P.~Schuster and N.~Toro,
  Phys.\ Rev.\  D {\bf 80}, 075018 (2009)
  [arXiv:0906.0580 [hep-ph]].
  
\bibitem{ADM}
  D.~E.~Kaplan, M.~A.~Luty and K.~M.~Zurek,
  Phys.\ Rev.\  D {\bf 79}, 115016 (2009)
  [arXiv:0901.4117 [hep-ph]].
  
\bibitem{QDM}
  G.~D.~Kribs, T.~S.~Roy, J.~Terning and K.~M.~Zurek,
  arXiv:0909.2034 [hep-ph].
  
\bibitem{D0}
  V.~M.~Abazov {\it et al.}  [D0 Collaboration],
  Phys.\ Rev.\ Lett.\  {\bf 103}, 081802 (2009)
  [arXiv:0905.1478 [hep-ex]].

  
\end{thebibliography}

\end{document}